\documentclass[12pt]{article}
\usepackage{graphics}
\usepackage{epsfig,amsmath,amssymb}
\input epsf

\begin{document}
\thispagestyle{empty}
\begin{center}
{\Large\bf{The statistical parton distributions:\\
\vskip 0.3cm
status and prospects}} 
\vskip1.4cm
{\bf Claude Bourrely and Jacques Soffer}  
\vskip 0.3cm
Centre de Physique Th\'eorique, UMR 6207 \footnote{UMR 6207 - Unit\'e Mixte 
de Recherche du CNRS et des Universit\'es Aix-Marseille I,
Aix-Marseille II et de l'Universit\'e du Sud Toulon-Var - Laboratoire 
affili\'e à la FRUMAM},\\
CNRS-Luminy, Case 907\\
F-13288 Marseille Cedex 9 - France \\ 
\vskip 0.5cm
{\bf Franco Buccella}
\vskip 0.3cm
Dipartimento di Scienze Fisiche, Universit\`a di Napoli,\\
Via Cintia, I-80126, Napoli
and INFN, Sezione di Napoli, Italy
\end{center}
\vskip 1.5cm
\begin{center}
{\bf Abstract}
\end{center}
New experimental results on polarized structure functions, cross sections for
$e^{\pm}p$ neutral and charge current reactions and $\nu$ ($\bar{\nu}$)
charge current on isoscalar targets are compared with predictions using the 
statistical parton distributions, which were previously determined.
New data on cross sections for Drell-Yan processes, single jet in $p\bar{p}$ 
collisions and inclusive $\pi^0$ production in $pp$ collisions 
are also compared with predictions from this theoretical approach.
The good agreement which we find with all these tests against experiment, 
strenghtens our opinion on the relevance of the role of quantum
statistics for parton distributions. We will also discuss the prospects of this
physical framework.\\

\vskip 0.5cm 

\vskip 0.5cm
\noindent CPT-2004/P.090

\noindent UNIV. NAPLES DSF 032/04
\newpage
\section{Introduction}
Deep-inelastic scattering (DIS) of leptons and nucleons is, so far, our
main source of information to study the internal nucleon structure, in terms
of parton distributions. Three years ago we proposed \cite{BBS1}  
to construct, in a unique way, the unpolarized and the polarized parton 
distributions, using a simple procedure, inspired by a quantum statistical 
picture of the nucleon, in terms of Fermi-Dirac and Bose-Einstein functions.
An important feature of this new approach lies into the fact that the chiral
properties of perturbative quantum chromodynamics (QCD), lead to strong 
relations between quark and antiquark distributions. 
As a consequence the determination of the best known unpolarized
light quarks ($u,d$) distributions and their corresponding polarized ones 
($\Delta u, \Delta d$), allows to predict the light antiquarks distributions 
($\bar u,\bar d,\Delta \bar u,\Delta \bar d$). 
Therefore our approach has a strong predictive power, in particular, 
the flavor-asymmetric light sea, {\it i.e.} $\bar d > \bar u$, which can be
understood in terms of the Pauli exclusion principle, based on the fact 
that the proton contains two $u$ quarks and only one $d$ quark \cite{Pauli}.
It is also natural to anticipate that the signs of 
$\Delta \bar u$ and $\Delta \bar d$ are the same as $\Delta u$ and $\Delta d$, 
respectively. 
One more relevant point to recall, is that all these parton distributions were 
determined in terms of only {\it eight} free parameters, which is indeed 
remarkable.

More recently we compared \cite{BBS2} our predictions with some new 
unpolarized and polarized DIS measurements obtained at DESY, SLAC and
Jefferson Lab. and they turned out to be rather satisfactory. Therefore, in
order to strengthen the relevance of this physical picture, we carry on the 
comparison with data from a much broader set of processes, including new DIS
results and also hadronic cross sections.

The paper is organized as follows. In the next section, we review the main 
points of our approach for the construction of the statistical
parton distributions and we recall their explicit expressions.
In section 3, we discuss in more details the predictive power of our approach
in connection with some simple mathematical properties of the Fermi-Dirac 
expressions and the numerical values we found for the free parameters. 
It allows us to clarify the $x$ behavior of the quark distributions, where it 
is known from the data, but also to foresee some specific behaviors, in so far
unexplored regions, for example in the high $x$ domain.
In section 4, we consider $e^{\pm}p$ neutral and charged current reactions,
whereas section 5 deals with $\nu(\bar{\nu})p$ charged current reactions.
Section 6 concerns  Drell-Yan processes, while section 7 deals with inclusive
single-jet production in $p\bar{p}$ collisions and inclusive $\pi^0$
production in $pp$ collisions. We give our final remarks and conclusions 
in the last section.

\section{The quantum statistical parton distributions}

The light quarks $q=u,d$ of helicity $h=\pm$, at the input energy scale 
$Q_0^2=4\mbox{GeV}^2$, are given by the sum of two terms \cite{BBS1},
a quasi Fermi-Dirac function and a helicity independent diffractive
contribution, common to all light quarks
\begin{equation}
xq^{h}(x,Q_0^2)= \frac{A X_{0q}^h x^b}{\exp[(x-X_{0q}^h)/{\bar x}]+1} +
\frac{{\tilde A} x^{\tilde b}}{\exp(x/{\bar x})+1}~.
\label{1}
\end{equation}
Here $X^{h}_{0q}$ is a constant, which plays the role of the 
{\it thermodynamical potential} of the quark $q^h$ and $\bar{x}$ is the 
{\it universal temperature}, which is the same for all partons.
We recall that from the chiral structure of QCD, we have two important 
properties, allowing to relate quark and antiquark distributions and to
restrict the gluon distribution \cite{BSa,BSb,Bha}:

- The potential of a quark $q^{h}$ of helicity {\it h} is opposite to the
potential of the corresponding antiquark $\bar q^{-h}$ of helicity {\it -h}
\begin{equation}
X_{0q}^h=-X_{0\bar q}^{-h}~.
\label{2}
\end{equation}

- The potential of the gluon $G$ is zero
\begin{equation}
X_{0G}=0~.
\label{3}
\end{equation}
Therefore similarly to Eq.~(\ref{1}), we have for the light antiquarks
\begin{equation}
x\bar q^{h}(x,Q_0^2)= \frac{\bar A (X_{0q}^{-h})^{-1}
x^{2b}}{\exp[(x+X_{0q}^{-h})/{\bar x}]+1}
+\frac{{\tilde A}x^{\tilde b}}{\exp(x/{\bar x})+1}~.
\label{4}
\end{equation}
Here we take $2b$ for the power of $x$ and not $b$ as for quarks, 
an assumption which was discussed and partly justified in Ref.~\cite{BBS1}.\\
Concerning the unpolarized gluon distribution, we use a quasi Bose-Einstein 
function, with zero potential
\begin{equation}
xG(x,Q_0^2)=\frac{A_Gx^{b_G}}{\exp(x/{\bar x})-1}~.
\label{5}
\end{equation}
This choice is consistent with the idea that hadrons, in the DIS regime, 
are black body cavities for the color fields. 
It is also reasonable to assume that for very small {\it x}, $xG(x,Q_0^2)$ 
has the same behavior as the diffractive contribution of the quark and
antiquark distributions in Eqs.~(\ref{1}) and (\ref{4}), so we will take 
$b_G=1+\tilde b$. We also need to specify the polarized gluon distribution and 
we take 
\begin{equation}
x \Delta G(x,Q_0^2) = 0~,
\label{6}
\end{equation}
assuming a zero polarized gluon distribution at the input energy scale $Q_0^2$
.\\
For the strange quarks and antiquarks, {\it s} and  $\bar s $, given our poor
knowledge on their unpolarized and polarized distributions, we take 
\footnote{A strangeness asymmetry, $s(x) \neq \bar s(x)$, can be also 
obtained in the statistical approach \cite{BBS5}}
\begin{equation}
xs(x,Q_0^2)=x \bar s(x,Q_0^2)= \frac{1}{4}[x \bar u (x,Q_0^2) + x \bar
d(x,Q_0^2)]~,
\label{7}
\end{equation}
and
\begin{equation}
x\Delta s(x,Q_0^2)=x \Delta \bar s(x,Q_0^2)= \frac{1}{3}[x \Delta \bar d
(x,Q_0^2) -x \Delta \bar u(x,Q_0^2)]~.
\label{8}
\end{equation}
This particular choice gives rise to a large negative $\Delta s(x,Q_0^2)$.
Both unpolarized and polarized distributions for the heavy quarks 
{\it c, b, t}, are set to zero at $Q_0^2=4\mbox{GeV}^2$.

With the above assumptions, we note that the heavy quarks do not introduce any
free parameters, likewise the gluons, since the normalization constant 
$A_G$ in Eq.~(\ref{5}) is determined from the momentum sum rule. 
Among the parameters introduced so far in Eqs.~(\ref{1}) and (\ref{4}),
$A$ and $\bar{A}$ are fixed by the two conditions $u-\bar{u}=2$, $d-\bar{d}=1$.
Clearly these valence quark conditions are independent of 
$\tilde b$ and $\tilde A$, since the diffractive contribution cancels out.
Therefore the light quarks require only {\it eight} free parameters, the 
{\it four} potentials $X^+_{0u}$, $X^-_{0u}$, $X^+_{0d}$, $X^-_{0d}$, 
{\it one} universal temperature $\bar x$, $b$, $\tilde b$ and $\tilde A$.

From well established features of the $u$ and $d$ quark distributions
extracted from DIS data, we anticipate some simple relations 
between the potentials:

- $u(x)$ dominates over $d(x)$, therefore one expects 
$X^{+}_{0u} + X^{-}_{0u} > X^{+}_{0d} + X^{-}_{0d}$

- $\Delta u(x) > 0$, therefore $X_{0u}^+ > X_{0u}^-$

- $\Delta d(x) < 0$, therefore $X_{0d}^- > X_{0d}^+$~.

So $X_{0u}^+$ should be the largest thermodynamical potential and
$X_{0d}^+$ the smallest one.
In fact, as we will see below, we have the following ordering
\begin{equation}
X_{0u}^+ > X_{0d}^- \sim  X_{0u}^- > X_{0d}^+ ~.
\label{9}
\end{equation} 
This ordering leads immediately to some important consequences for quarks and 
antiquarks. 

\begin{figure}[htb]
\begin{center}
\leavevmode {\epsfysize=13.0cm \epsffile{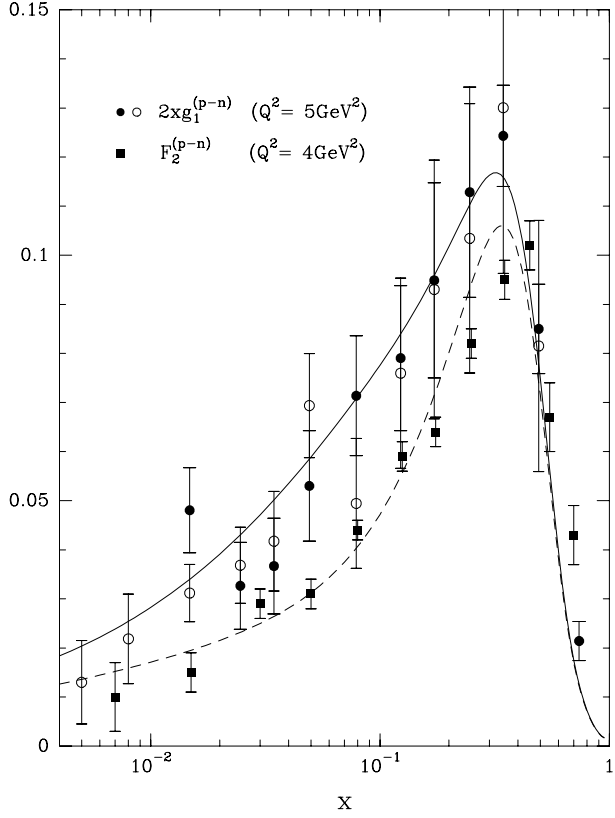}}
\end{center}
\vspace*{-25mm}
\caption[*]{\baselineskip 1pt
The isovector structure functions $2xg_1^{(p-n)}(x)$ (solid line from our 
statistical parton distributions) and $F_2^{(p-n)}(x)$
(dashed line from our statistical distributions).
Data are from NMC \cite{NMC}, SMC \cite{SMC}, SLAC \cite{slace155a}, 
JLab \cite{JLab04} and COMPASS \cite{COM05}.}
\label{fi:isovec}
\vspace*{-1.5ex}
\end{figure}

First, the fact that $X_{0d}^- \sim  X_{0u}^-$, indicated in Eq.~(\ref{9}), 
leads to
\begin{equation}
u^-(x,Q_0^2) \lesssim d^-(x,Q_0^2)~,
\label{10}
\end{equation}
which implies from our procedure to construct antiquark from quark 
distributions,
\begin{equation}
\bar u^+(x,Q_0^2) \gtrsim \bar d^+(x,Q_0^2)~.
\label{11}
\end{equation}
These two important approximate relations were already obtained in 
Ref.~\cite{BBS1}, by observing in the data, the similarity in shape 
of the isovector structure functions
$2xg_1^{(p-n)}(x)$ and $F_2^{(p-n)}(x)$, at the initial energy scale, 
as illustrated in Fig.~\ref{fi:isovec} 
~\footnote{Notice that it differs from Fig.~1 in Ref.~\cite{BBS1}, where
we put incorrect scales, both on the vertical and the horizontal axes}. 
For $2xg_1^{(p-n)}(x)$ the black circles are obtained by combining SLAC 
\cite{slace155a} and JLab \cite{JLab04} data.
The white circles, which extend down to the very low $x$ region, include the 
recent deuteron data from COMPASS \cite{COM05} combined with the 
proton data from SMC \cite{SMC}, at the measured $Q^2$ values of these two 
experiments \footnote{We have not included some corrections due to difference
of the beam energies of COMPASS and SMC}. The agreement with the curve of the
statistical model is improved in this later case. The + helicity components
disappear in the difference $2xg_1^{(p-n)}(x)-F_2^{(p-n)}(x)$. Since this 
difference
is mainly non-zero for $0.01 < x < 0.3$, it is due to the contributions of
$\bar u^-$ and $\bar d^-$ (see Ref.~\cite{BBS1}).

Second, the ordering in Eq.~(\ref{9}) implies the following properties for 
antiquarks, namely:

i) $\bar d(x) > \bar u(x)$, the flavor symmetry breaking which also follows
from the Pauli exclusion principle, as recalled above. 
This was already confirmed by the violation of the Gottfried sum rule 
\cite{Gott,NMC}.

ii) $\Delta \bar u(x) > 0$ and  $\Delta \bar d(x) < 0$, which have not been
established yet, given the lack of precision of the polarized semi-inclusive 
DIS data, as we will see below. One expects an accurate 
determination of these distributions from the measurement of helicity 
asymmetries for weak boson production in polarized $pp$ collisions at RHIC-BNL 
\cite{BSSW}, which will allow this flavor separation.

By performing a next-to-leading order QCD evolution of these parton 
distributions, we were able to obtain in Ref.~\cite{BBS1}, a good description 
of a large set of very precise data on the following unpolarized 
and polarized DIS structure functions 
$F_2^{p, d, n}(x,Q^2), xF_3^{\nu N}(x,Q^2)$
and $g_1^{p, d, n}(x,Q^2)$, in a broad range of $x$ and $Q^2$,
in correspondance with the {\it eight} free parameters :
\begin{equation}
X_{0u}^{+} = 0.46128,~ X_{0u}^{-} = 0.29766,~X_{0d}^{-} = 0.30174,~X_{0d}^{+} 
= 0.22775 \, ,
\label{12}
\end{equation}
\begin{equation}
\bar x =0.09907,~ b = 0.40962,~\tilde b = -0.25347,~\tilde A =0.08318\, ,
\label{13}
\end{equation}
and three additional parameters, which are fixed by normalization conditions
\begin{equation}
A = 1.74938,~\bar A = 1.90801,~A_G = 14.27535 ~,
\label{14}
\end{equation}
as explained above. Note that the numerical values of the four potentials are 
in agreement with the ordering in Eq.~(\ref{9}), as expected, and all
the free parameters in Eqs.~(\ref{12}, \ref{13}) have been determined rather 
precisely, with an error of the order of one percent.

\section{The predicting power of the statistical parton distributions}

\begin{figure}[t]
\begin{center}
  \begin{minipage}{6.5cm}
  \epsfig{figure=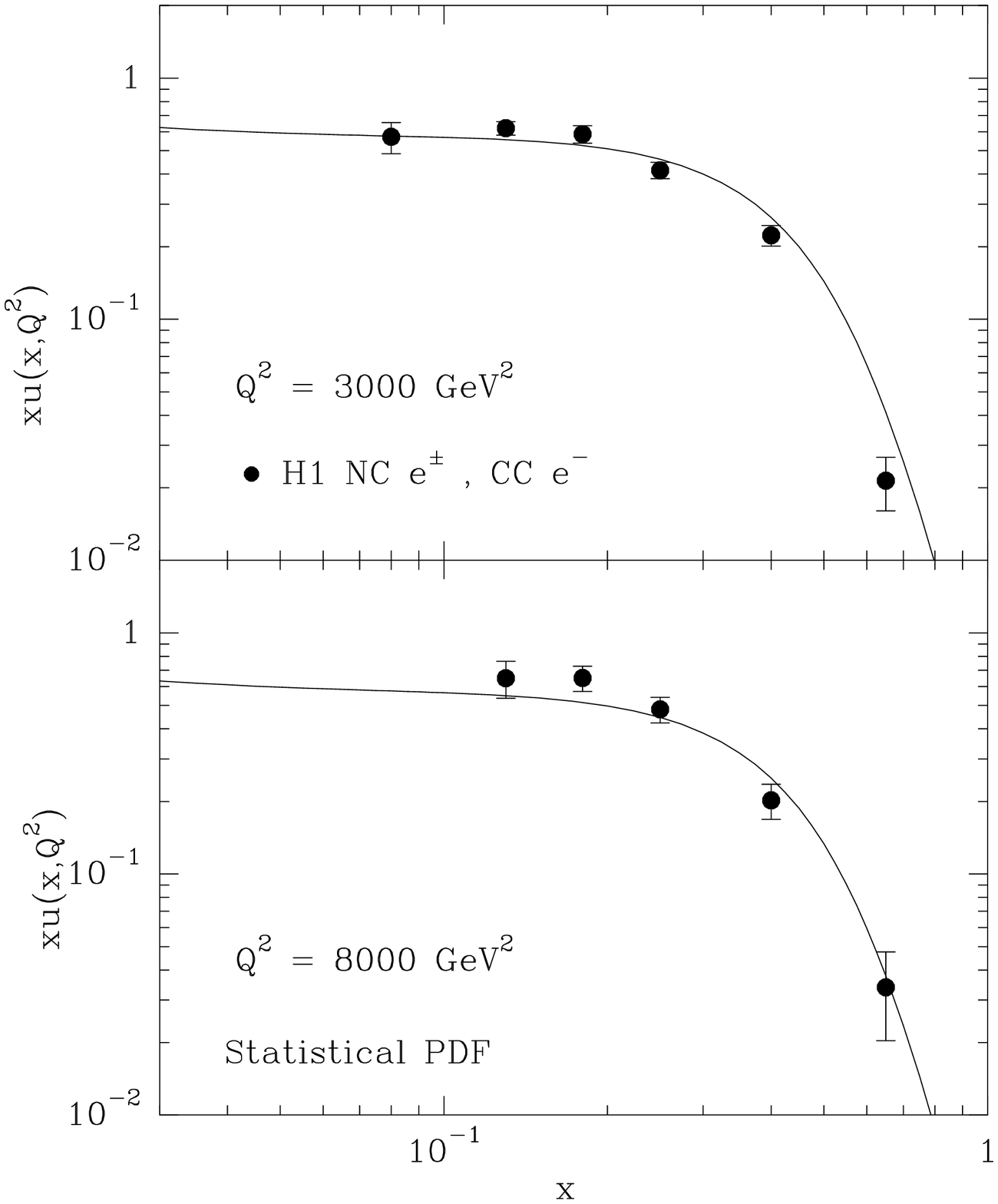,width=6.5cm}
  \end{minipage}
    \begin{minipage}{6.5cm}
  \epsfig{figure=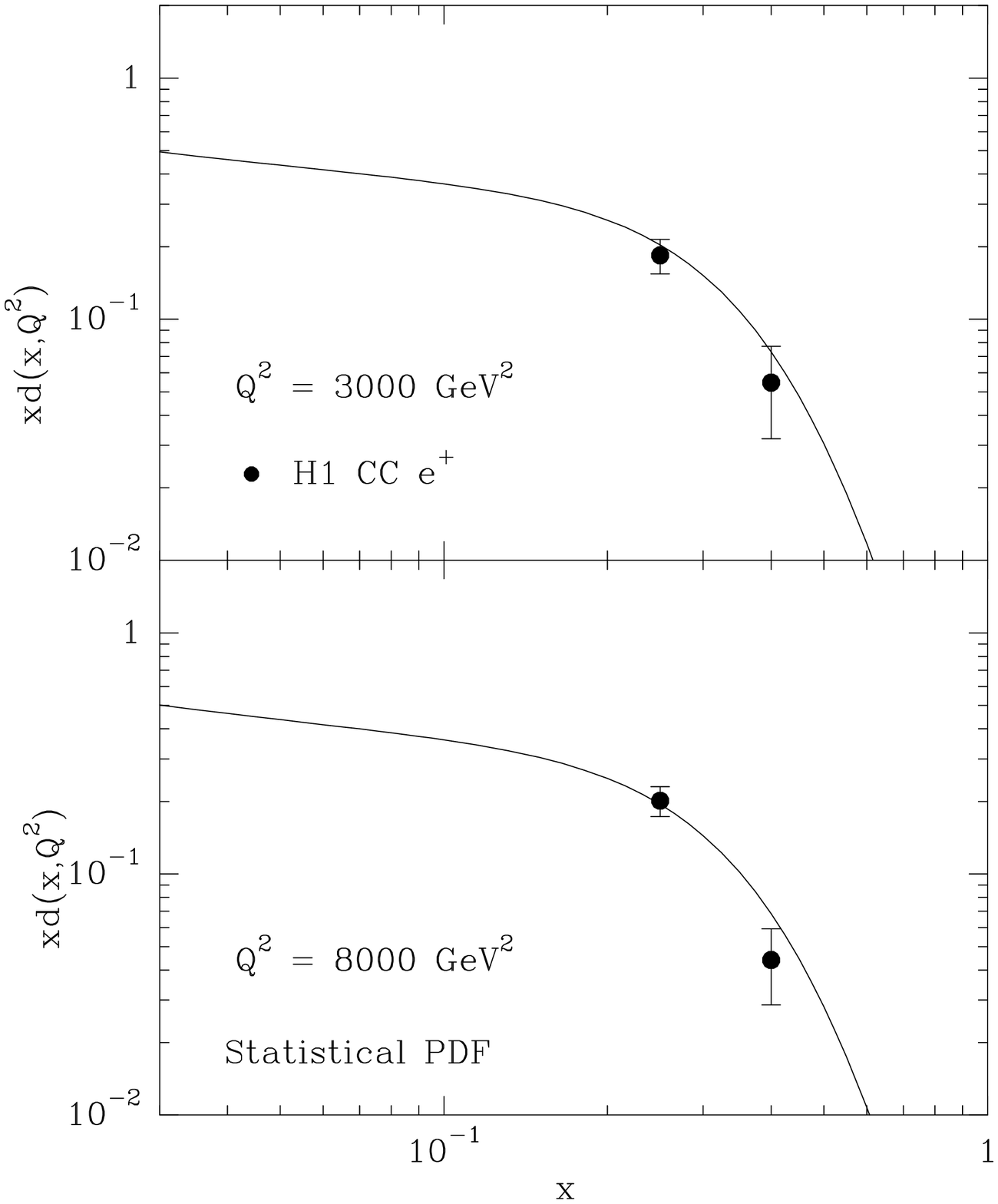,width=6.5cm}
    \end{minipage}
\end{center}
  \vspace*{-10mm}
\caption{
Statistical quark distributions $xu(x,Q^2),~xd(x,Q^2)$ as a function of $x$ for
$Q^2 = 3000,~8000\mbox{GeV}^2$. Data from H1 \cite{h103}.}
\label{fi:qvsx}
\vspace*{-1.5ex}
\end{figure}

\begin{figure}[t]
\begin{center}
  \begin{minipage}{6.5cm}
  \epsfig{figure=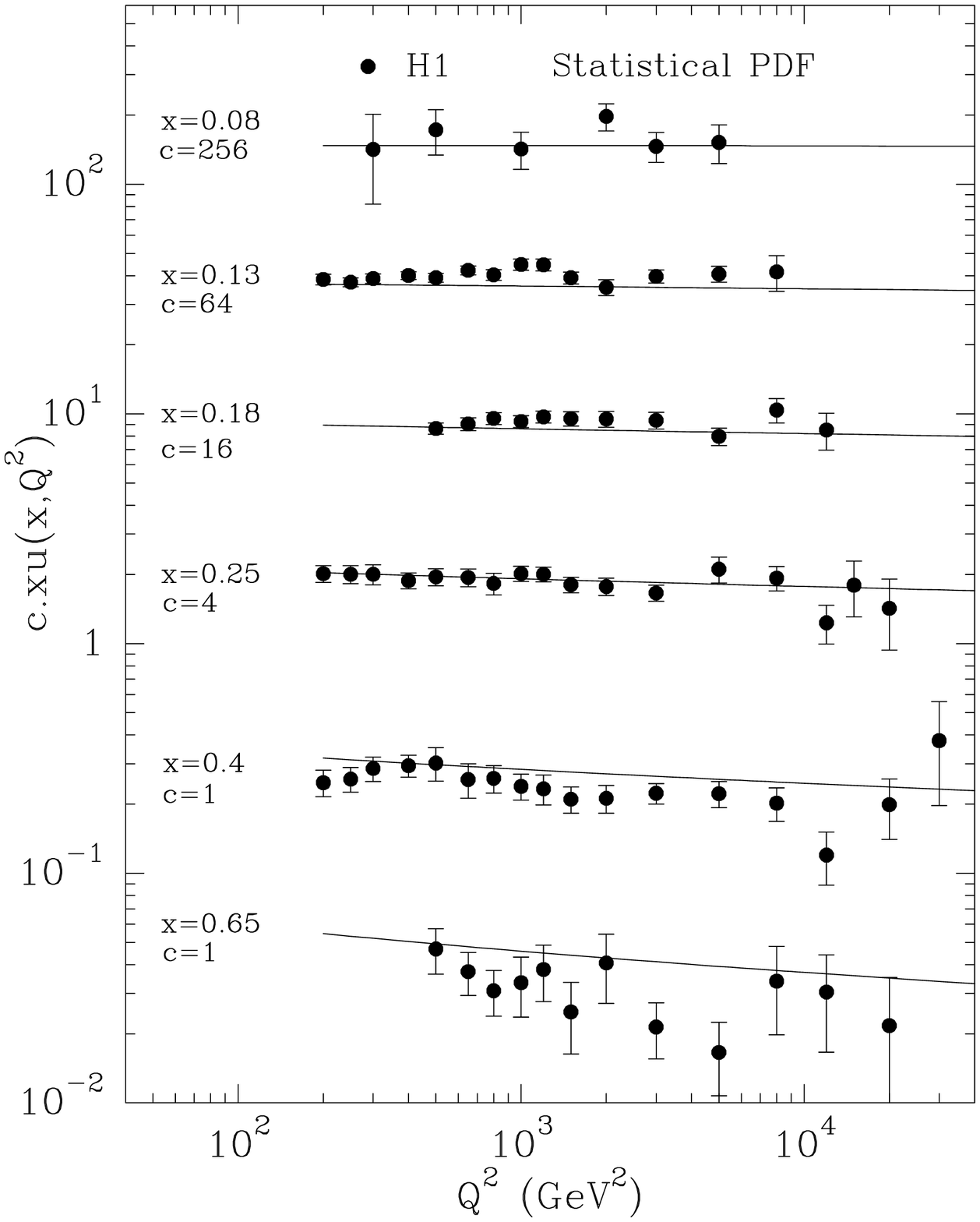,width=6.5cm}
  \end{minipage}
    \begin{minipage}{6.5cm}
  \epsfig{figure=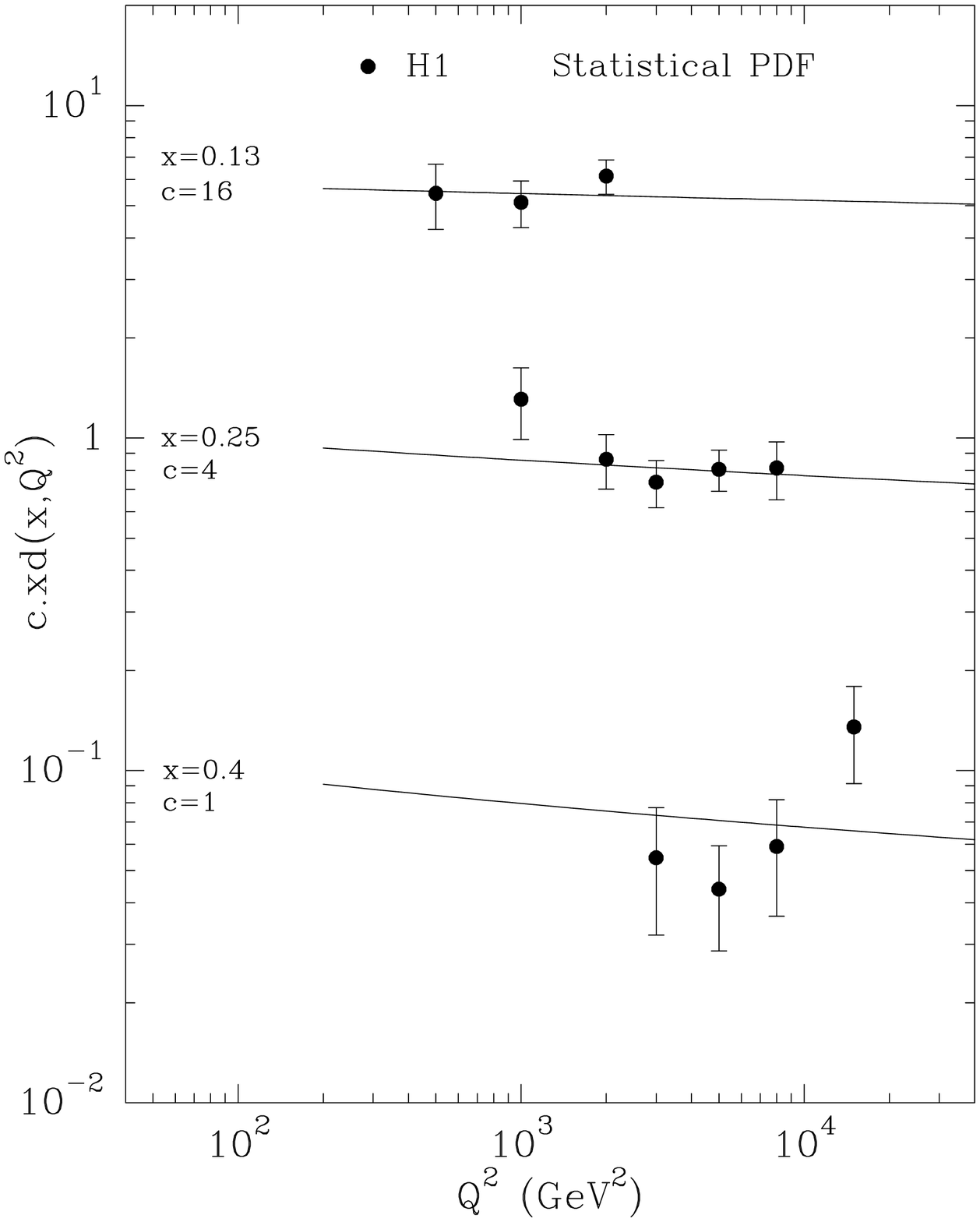,width=6.5cm}
    \end{minipage}
\end{center}
  \vspace*{-10mm}
\caption{
Statistical quark distributions $c\cdot xu(x,Q^2),~c\cdot xd(x,Q^2)$ 
as a function of $Q^2$ for fixed $x$ bins. Data from H1  \cite{h103}.}
\label{fi:qvsq2}
\vspace*{-1.5ex}
\end{figure}

\begin{figure}[htb]
\begin{center}
\leavevmode {\epsfysize=8.0cm \epsffile{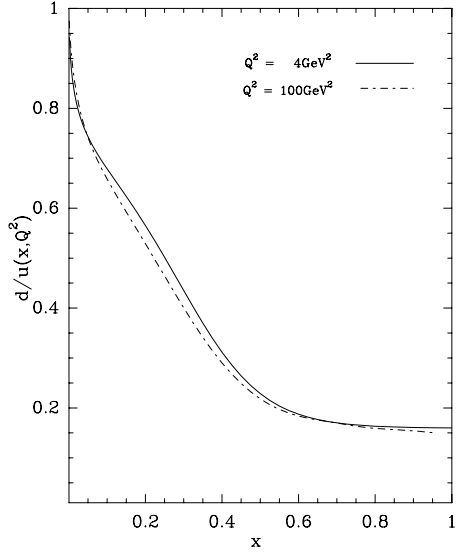}}
\end{center}
\vspace*{-10mm}
\caption[*]{\baselineskip 1pt
The quarks ratio $d/u$ as function of $x$ for $Q^2 = 4\mbox{GeV}^2$ 
(solid line) and $Q^2 =100\mbox{GeV}^2$ (dashed-dotted line).}
\label{fi:doveru}
\vspace*{-3.5ex}
\end{figure}

We now try to relate the $x$ dependence of the quark (antiquark) distributions
to their specific expressions given in Eqs.~(\ref{1}) and (\ref{4}) and 
to study the role of the different free parameters involved, according to 
their numerical values obtained in Ref.~\cite{BBS1}.
First, it is useful to note that, given the small value of $\tilde A$ 
(see Eq.~(\ref{13})), the diffractive contribution is less than $10^{-2}$
or so, for $x\geq 0.1$, but it dominates in the very low $x$ region,
when $x<<\bar x$, since $\tilde b <0$. Therefore the strong change
of slope of $xu(x)$ and $xd(x)$ at high $x$ (at the input scale 
$Q^2_0$ and above), is related to the values of the corresponding 
potentials and is larger for $u$ than for $d$, because of 
the ordering in Eq.~(\ref{9}). 
This is indeed what we observe in Fig.~\ref{fi:qvsx}, at some rather high 
$Q^2$ values.
This feature is not spoilt by the $Q^2$ evolution, which is also well 
described by the statistical quark distributions as shown 
in Fig.~\ref{fi:qvsq2}, where we compare with H1 data. Another interesting 
point concerns the behavior of the ratio $d(x)/u(x)$, 
which depends on the mathematical properties of the ratio of two Fermi-Dirac 
factors, outside the region dominated by the diffractive contribution. 
So for $x>0.1$, this ratio is expected to decrease faster for 
$X_{0d}^+ - \bar x < x < X_{0u}^+ + \bar x$ and then above, for 
$x > 0.6$ it flattens out.
This change of slope is clearly visible in Fig.~\ref{fi:doveru}, with a very 
little $Q^2$ dependence. Note that our prediction for the large $x$ behavior,
differs from most of the current literature, namely $d(x)/u(x) \to 0$
for $x \to 1$, but we find $d(x)/u(x) \to 0.16$ near the value $1/5$,
a prediction originally formulated in Ref.~\cite{FJ}.
This is a very challenging question, since the very high $x$ region remains
poorly known, as shown in Fig.~\ref{fi:qvsx} and Fig.~\ref{fi:qvsq2}.
The typical behaviour of the Fermi-Dirac functions, falling down
exponentially above the thermodynamical potential, which shows up in
Fig.~\ref{fi:isovec}, complies well with the fast change in the slope of
$g_1^p(x)$ at high $x$, as shown in Fig.~\ref{fi:g1pe143}.

\begin{figure}[htb]
\begin{center}
\leavevmode {\epsfysize=9.5cm \epsffile{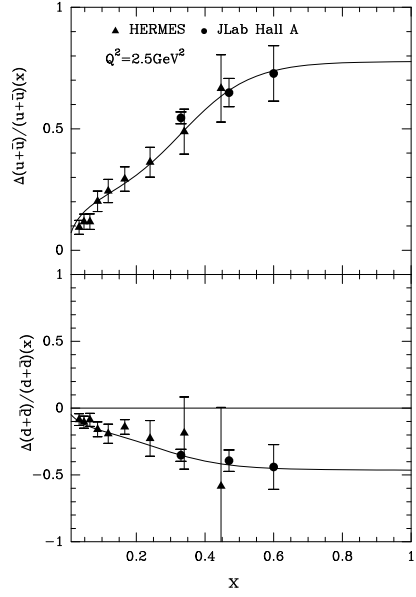}}
\end{center}
\vspace*{-15mm}
\caption[*]{\baselineskip 1pt
Ratios $(\Delta u + \Delta \bar u)/(u + \bar u)$ and 
$(\Delta d + \Delta \bar d)/(d + \bar d)$  as a function of $x$.
Data from Hermes for $Q^2 = 2.5\mbox{GeV}^2$ \cite{herm99} and
a JLab experiment \cite{JLab04a}. The curves are predictions from the 
statistical approach.}
\label{fi:rapdelq}
\vspace*{-2.5ex}
\end{figure}

\begin{figure}
\begin{center}
\leavevmode {\epsfysize=8.5cm \epsffile{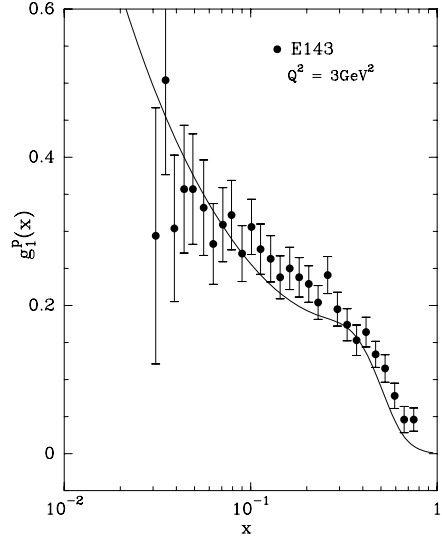}}
\end{center}
\vspace*{-15mm}
\caption[*]{\baselineskip 1pt
$g_1^p(x,Q^2)$ as a function of $x$ at fixed $Q^2 = 3\mbox{GeV}^2$
from the statistical approach.
Experimental data from SLAC E143 \cite{slace143}.}
\label{fi:g1pe143}
\vspace*{-1.5ex}
\end{figure}

\begin{figure}[htb]
\begin{center}
\leavevmode {\epsfysize=8.5cm \epsffile{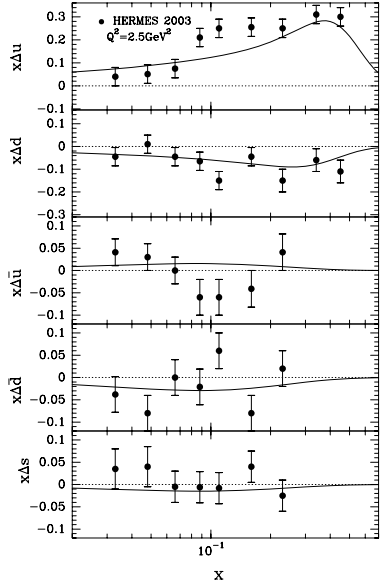}}
\end{center}
\vspace*{-5mm}
\caption[*]{\baselineskip 1pt
Quarks and antiquarks polarized parton distributions as a function of $x$
for $Q^2 = 2.5\mbox{GeV}^2$. Data from Hermes \cite{herm04}. The curves are 
predictions from the statistical approach.}
\label{fi:polpdf}
\vspace*{-1.5ex}
\end{figure}

\begin{figure}[htb]
\begin{center}
\leavevmode {\epsfysize=8.0cm \epsffile{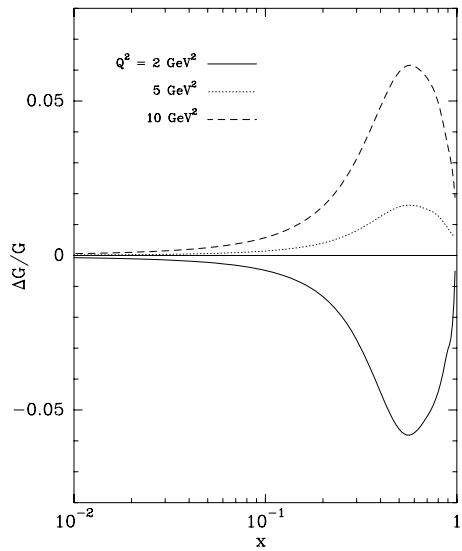}}
\end{center}
\vspace*{-10mm}
\caption[*]{\baselineskip 1pt
The ratio $\Delta G(x)/G(x)$ as a function of $x$, for 
$Q^2 = 2, 5~ \mbox{and}\\
10~\mbox{GeV}^2$. The curves are predictions from the statistical approach.}
\label{fi:delgsg}
\vspace*{-1.5ex}
\end{figure}

Analogous considerations can be made for the corresponding helicity 
distributions, whose best determinations are shown in Fig.~\ref{fi:rapdelq}.
By using a similar argument as above, the ratio $\Delta u(x)/u(x)$ 
is predicted to have a rather fast increase in the $x$ range 
$(X^-_{0u}-\bar{x},X^+_{0u}+\bar{x})$
and a smoother behaviour above, while $\Delta d(x)/d(x)$, which is negative,
has a fast decrease in the $x$ range $(X^+_{0d}-\bar{x},X^-_{0d}+\bar{x})$ 
and a smooth one above. This is exactly the trends displayed in 
Fig.~\ref{fi:rapdelq} and our predictions are in perfect agreement
with the accurate high $x$ data \footnote{ It is worth mentioning that the 
Jefferson Lab points for the $d$ quark are those of Ref.~\cite{JLab04a}, 
which have been moved down compared to those of Ref.~\cite{JLab04}}.
We note the behavior near $x=1$, another typical property of the statistical
approach, also at variance with predictions of the current literature. 
The fact that $\Delta u(x)$ is more concentrated in the higher $x$ region than
$\Delta d(x)$, accounts for the change of sign of $g^n_1(x)$, which becomes
positive for $x>0.5$, as first observed at Jefferson Lab \cite{JLab04}.\\
For the light antiquark distributions (see Eq.~(\ref{4})), it is clear that in
the very low $x$ region ($x<10^{-3}$) the ratio $\bar d(x)/\bar u(x)$ 
is $\sim 1$, since the diffractive contribution dominates 
\footnote{Obviously, this is also the case for the ratio $d(x)/u(x)$} and it 
is an increasing function of $x$ because the non diffractive term is larger 
for $\bar d(x)$ than for $\bar u(x)$. This natural expectation, 
$\bar d(x) \geq \bar u(x)$ from the statistical approach, was already
mentioned above and has been also confirmed by the E866/NuSea Drell-Yan
dilepton experiment \cite{E866}, up to $x=0.15$. For larger
$x$, although the errors are large, the data seem to drop off in disagreement 
with our predictions (see Fig. 16 in Ref.~\cite{BBS1}). 
This important point deserves further attention and we will come back to it in 
Section 6, when we will discuss Drell-Yan dilepton cross sections.
This is another challenging point, which needs to be clarified, for example 
with future measurements by the approved FNAL E906 experiment \cite{E906}, 
to higher $x$ values.\\
We now turn to the antiquark helicity distributions. Since we predict
$\Delta \bar u(x) > 0$ and  $\Delta \bar d(x) < 0$, the contribution of the 
antiquarks to the Bjorken sum rule (BSR) \cite{Bj} is in our case 0.022, 
at $Q^2=5 \mbox{GeV}^2$, which is not negligible.
The statistical model gives for the BSR the value 0.176, in excellent
agreement with the QCD prediction $0.182 \pm 0.005$ and with the
world data $0.176 \pm 0.003 \pm 0.07$~\cite{slace155a}. 
It is also interesting to remark that Eq.~(\ref{11}) implies
\begin{equation}
\Delta \bar u(x) - \Delta \bar d(x) \simeq \bar d(x) - \bar u(x) > 0 ~,
\label{15}
\end{equation}
so the flavor asymmetry of the light antiquark distributions is almost the 
same for the corresponding helicity distributions. Similarly, 
Eq.~(\ref{10}) implies
\begin{equation}
\Delta u(x) - \Delta d(x) \simeq  u(x) - d(x) > 0 ~.
\label{16}
\end{equation}
By combining Eqs.~(\ref{15}) and (\ref{16}), we find a very simple approximate
result for the BSR, namely $\sim 1/6$, a value compatible with the
numbers quoted above.
We also compare in Fig.~\ref{fi:polpdf} our predictions with an attempt from 
Hermes to isolate the different quark and antiquark helicity distributions. 
The poor quality of the data  does not allow to conclude on the signs of 
$\Delta \bar u(x)$ and $\Delta \bar d(x)$, which will have to wait for a
higher precision experiment.\\
Finally we are coming back to the polarized gluon distribution 
$\Delta G(x,Q^2)$, which was assumed to be zero at the input scale 
$Q^2_0=4\mbox{GeV}^2$ (see Eq.~(\ref{6})). It is interesting to note that
after evolution, it becomes negative for $Q^2 < Q^2_0$ and positive for 
$Q^2 > Q^2_0$. The results are displayed in Fig.~\ref{fi:delgsg} and are 
waiting for an improved experimental determination of this important 
distribution.

\newpage


\newpage

\clearpage
\newpage

\section{Inclusive neutral and charged current $e^{\pm}p$ cross sections}

\begin{figure}[t]
\begin{center}
  \begin{minipage}{6.5cm}
  \epsfig{figure=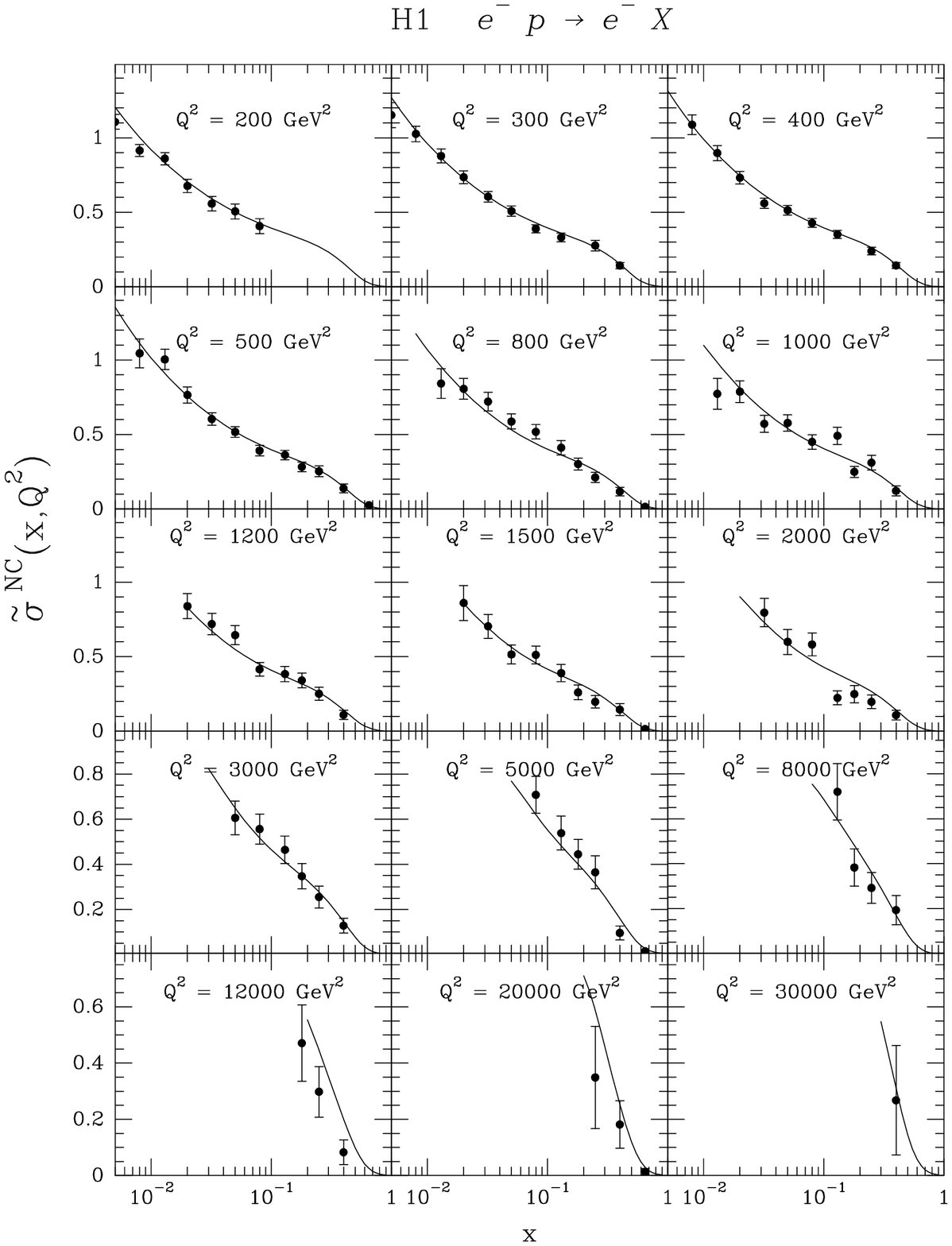,width=7.3cm}
  \end{minipage}
    \begin{minipage}{6.5cm}
  \epsfig{figure=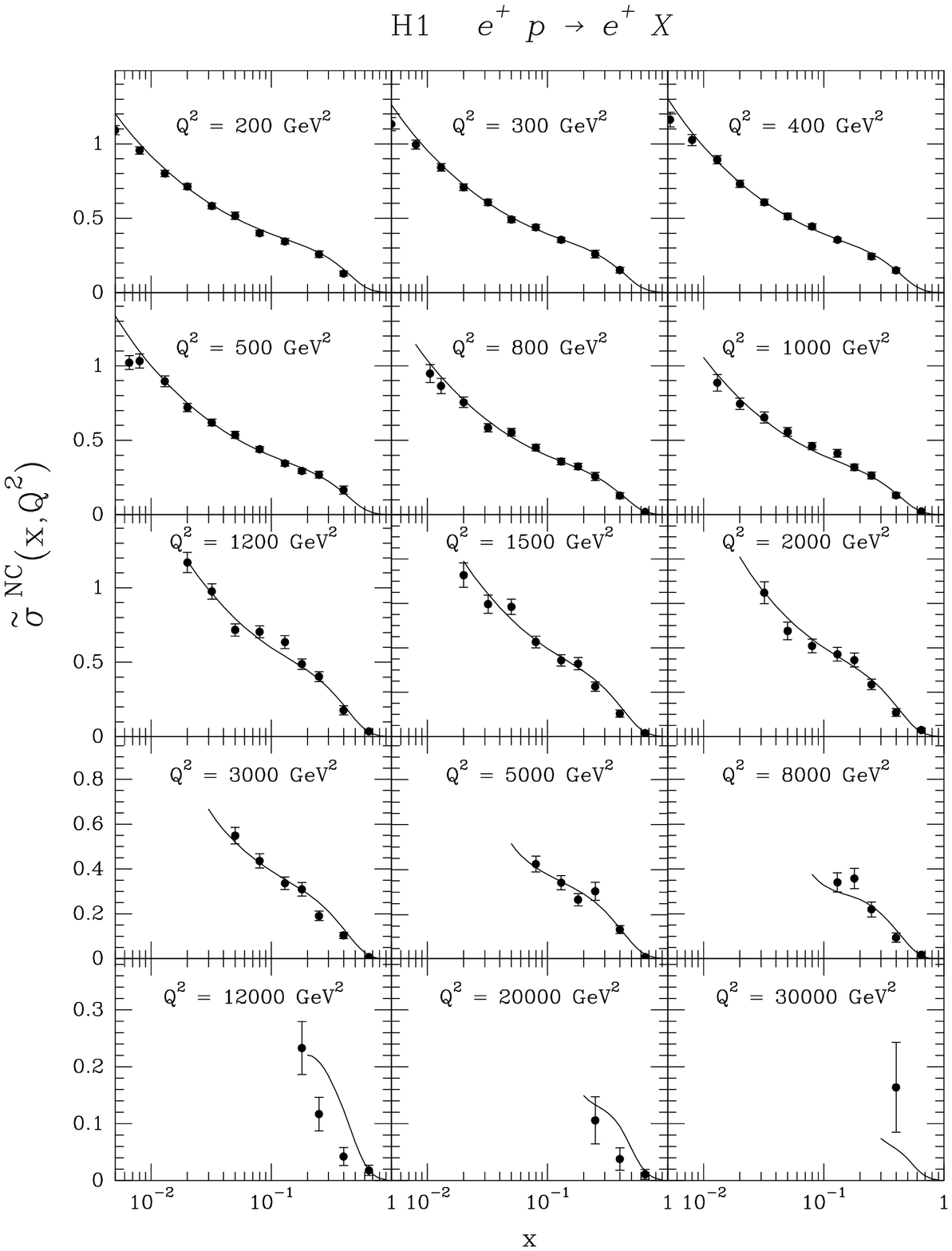,width=7.3cm}
    \end{minipage}\\
\end{center}
  \vspace*{-20mm}
\caption{
The reduced neutral current cross section $\tilde{\sigma}$,
as a function of $x$, for different fixed values of $Q^2$. 
Reaction $e^{-} p$ at $\sqrt{s} = 320\mbox{GeV}$, $e^{+} p$ at 
$\sqrt{s} = 319\mbox{GeV}$. Data from H1  \cite{h103,h100}.
The curves are predictions from the statistical approach.}
\label{fi:nche-p-vsx}
\vspace*{-7.5ex}
\begin{center}
  \begin{minipage}{6.5cm}
  \epsfig{figure=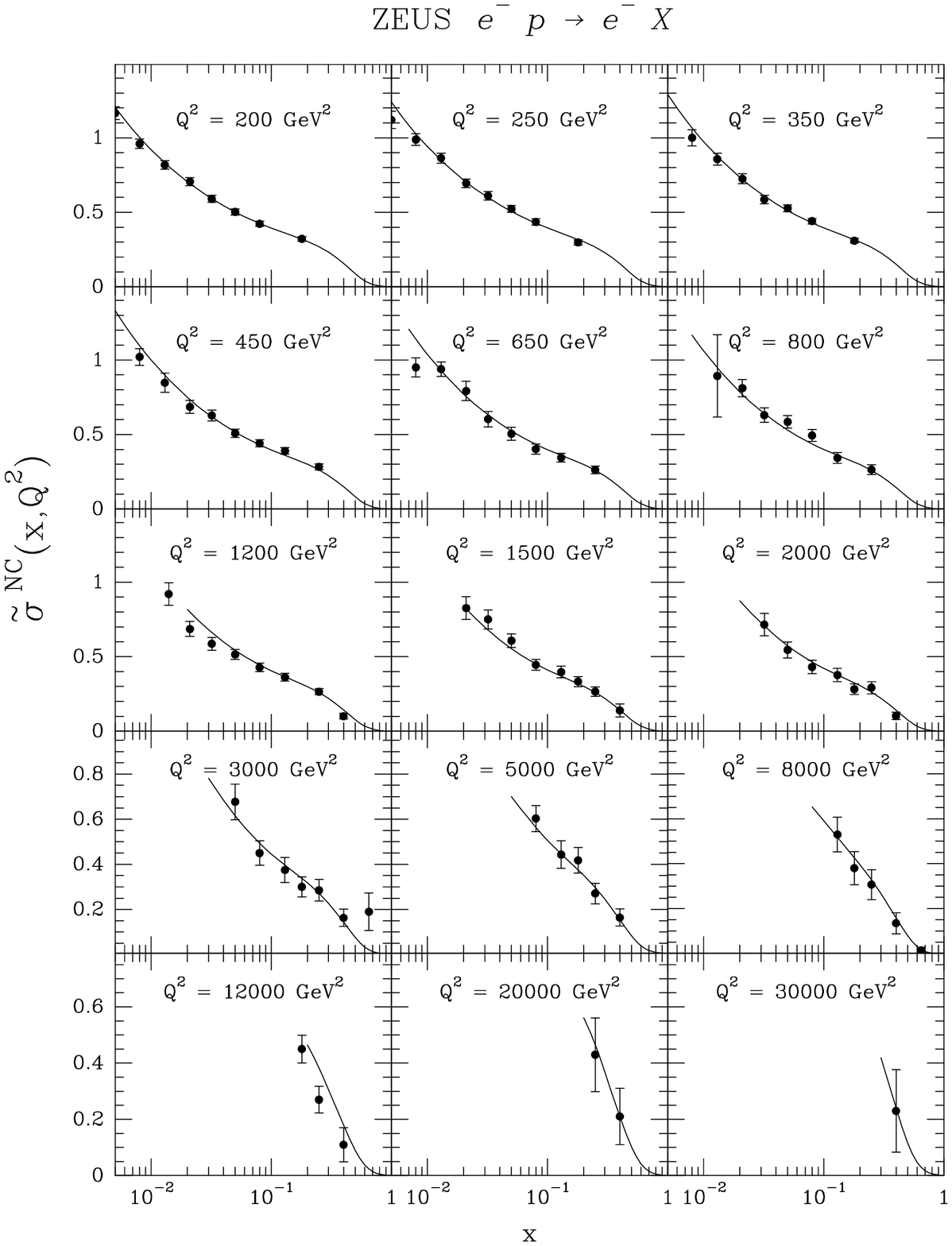,width=7.3cm}
  \end{minipage}
    \begin{minipage}{6.5cm}
  \epsfig{figure=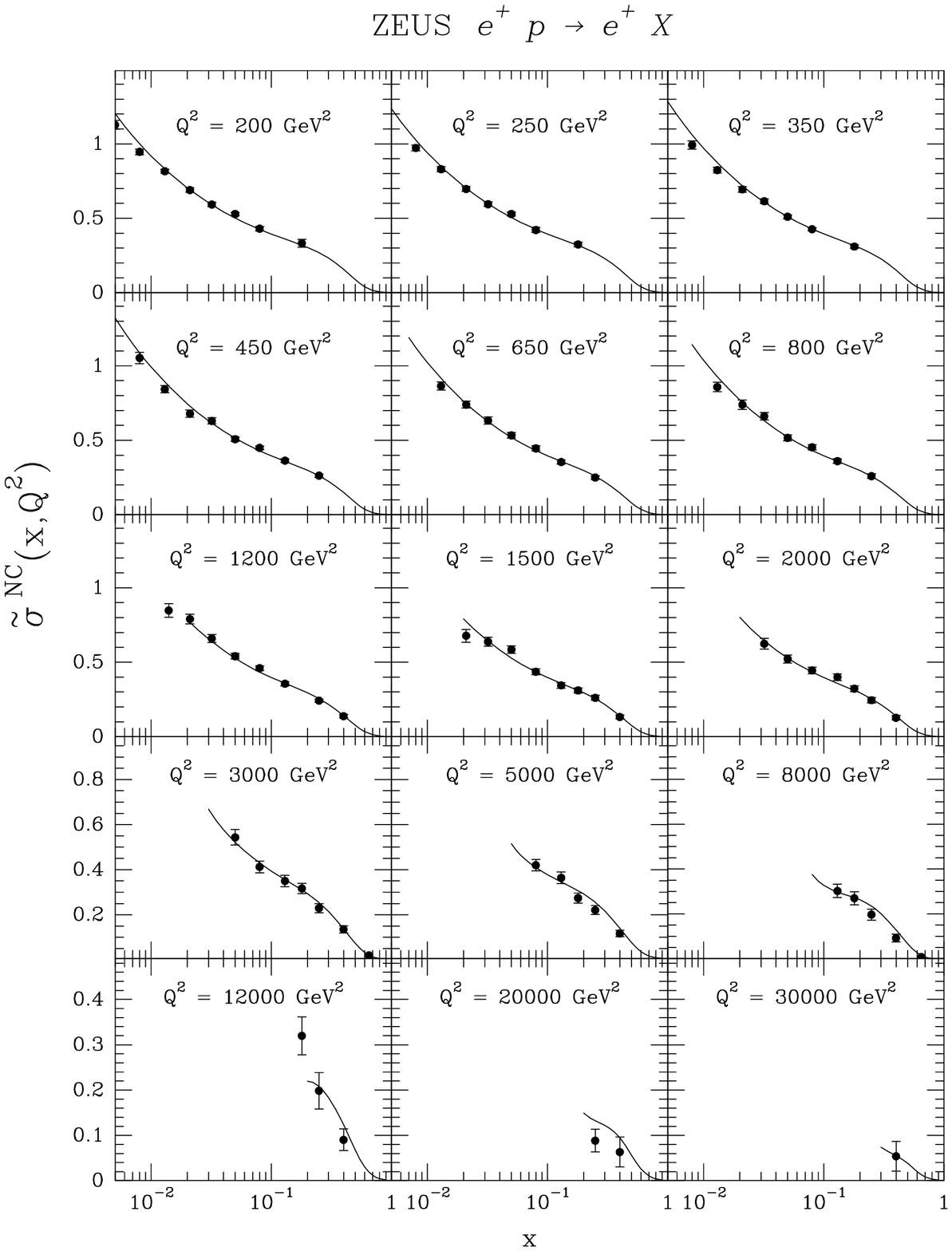,width=7.3cm}
    \end{minipage}
\end{center}
  \vspace*{-20mm}
\caption{
Same as Fig.~\ref{fi:nche-p-vsx}. Data from ZEUS \cite{zeus03}.}
\label{fi:ncze-p-vsq}
\vspace*{-5.5ex}
\end{figure}

The neutral current DIS processes have been measured at HERA in a kinematic 
region where both the $\gamma$ and the $Z$ exchanges must be considered.
The cross sections for neutral current can be written, at lowest order, as 
\cite{zhang}
\begin{equation}
\frac{d^2 \sigma^{\pm}_{NC}}{dx dQ^2} =
\frac{2\pi \alpha^2}{x Q^4}
\left[Y_+ \tilde F_2(x, Q^2) \mp Y_- x \tilde F_3(x, Q^2) - y^2
\tilde F_L(x, Q^2)\right]~,
\label{dcrossnc}
\end{equation}
where
\begin{equation}
\tilde F_2(x, Q^2) = F_2^{em} -v_e\chi_z(Q^2)G_2(x, Q^2) +
(a_e^2 + v_e^2)\chi^2_z(Q^2) H_2(x, Q^2)~,
\label{tildf2}
\end{equation}
\begin{equation}
x\tilde F_3(x, Q^2) = -a_e\chi_z(Q^2) xG_3(x, Q^2) + 2a_e v_e \chi^2_z(Q^2)
xH_3(x, Q^2)~.
\label{tildf3}
\end{equation}
The structure function $\tilde F_L(x, Q^2)$ is sizeable only at high $y$ and 
the other structure functions introduced above, have the following 
expressions in terms of the parton distributions
\begin{eqnarray}
\left[F_2^{em}, G_2, H_2\right](x, Q^2) &=&
\sum_f \left[Q^2_f, 2Q_f v_f, a^2_f + v^2_f\right]
\left(xq_f(x, Q^2) + x\bar q_f(x, Q^2)\right) , \nonumber \\
\left[xG_3, xH_3\right](x, Q^2) &=& \sum_f \left[2Q_f a_f, 2a_f v_f\right]
\left(xq_f(x, Q^2) - x \bar q_f(x, Q^2)\right) .
\label{fgh}
\end{eqnarray}
Here the kinematic variables are
$y = Q^2/xs$, $Y_{\pm} = 1 \pm (1-y)^2$, $\sqrt{s} = \sqrt{E_e E_p}$,
$E_e$ and $E_p$ are the electron (positron) and proton beam energies
respectively. Morever, $v_i$ and $a_i$ are the vector and axial-vector
weak coupling constants for the lepton $e$ and the quark $f$, respectively,
and $Q_f$ is the charge. The function $\chi_z(Q^2)$ is given by
\begin{equation}
\chi_z(Q^2) = \frac{1}{4\sin^2{\theta_W}\cos^2{\theta_W}}
\frac{Q^2}{Q^2 + M^2_Z}~,
\label{fchi2}
\end{equation}
where $\theta_W$ is the weak mixing angle and $M_Z$ is the $Z$-boson mass.
The reduced cross sections are defined as
\begin{equation}
\tilde \sigma^{\pm}_{NC}(x, Q^2) = 
\frac{Q^4 x}{Y_+ 2\pi \alpha^2}
\frac{d^2 \sigma^{\pm}_{NC}}{dx dQ^2}~.
\label{redncross}
\end{equation}
Our predictions are compared with H1 and ZEUS data in 
Figs.~\ref{fi:nche-p-vsx} and \ref{fi:ncze-p-vsq}, as a function of $x$, 
in a broad range of $Q^2$ values and the agreement is excellent.\\

\begin{figure}[t]
\begin{center}
  \begin{minipage}{6.5cm}
  \epsfig{figure=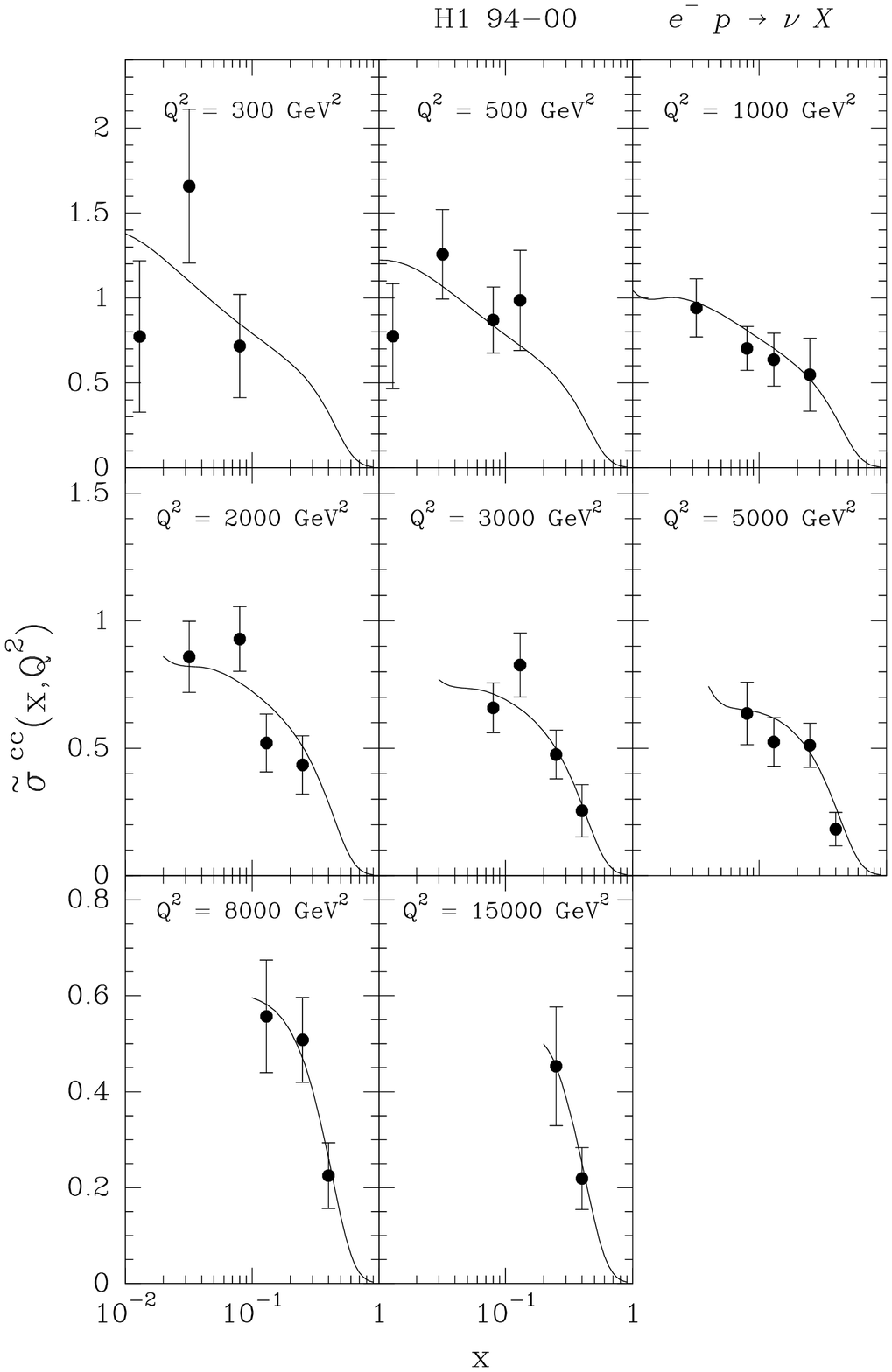,width=6.5cm}
  \end{minipage}
    \begin{minipage}{6.5cm}
  \epsfig{figure=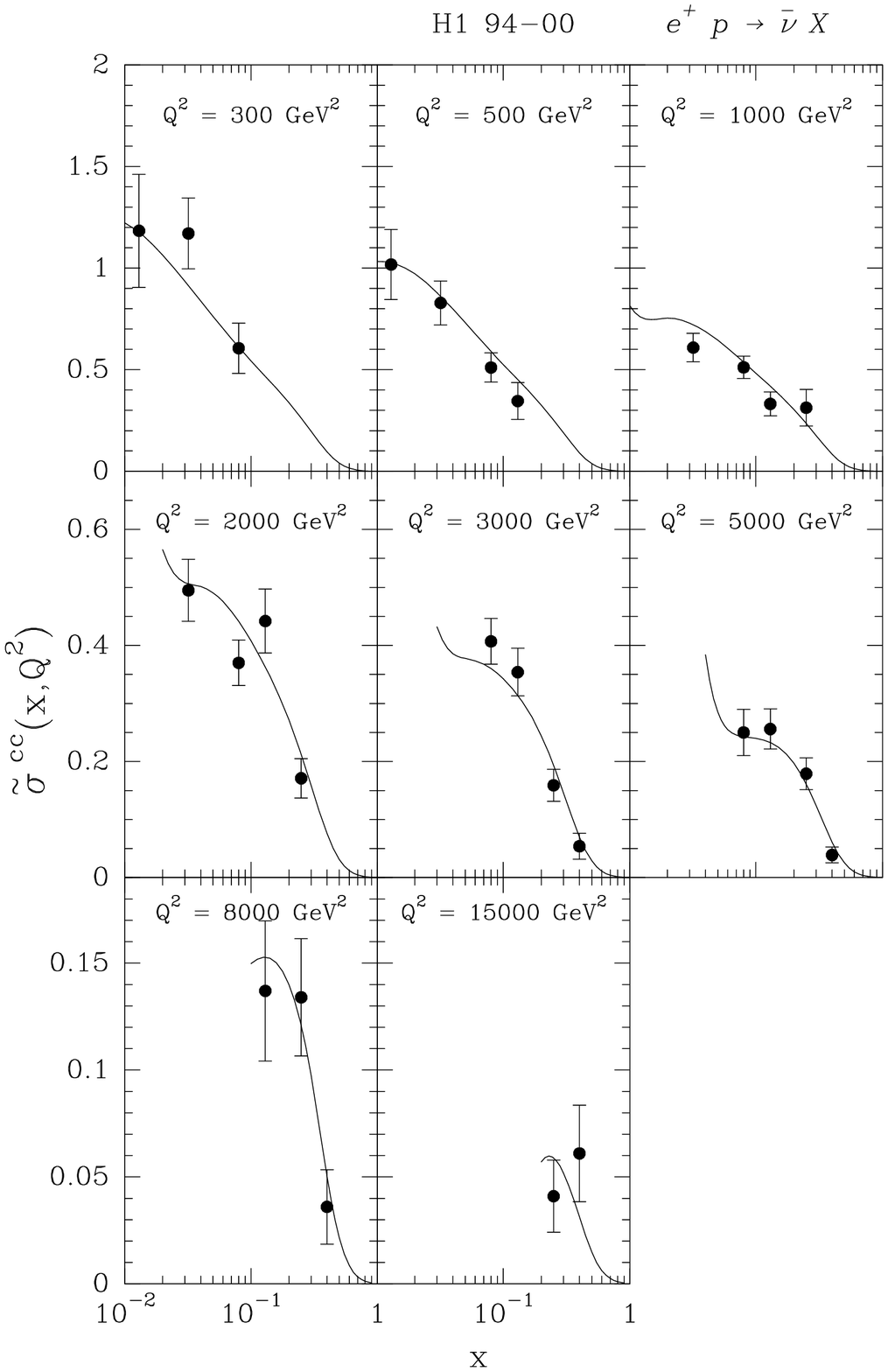,width=6.5cm}
    \end{minipage}\\
\end{center}
  \vspace*{-10mm}
\caption{
The reduced charged current cross section $\tilde{\sigma}$, in $e^{\pm} p$ 
reactions as a function of $x$, for different fixed values of $Q^2$. 
Data from H1  \cite{h103,h100}.}
\label{fi:he-p-vsx}
\vspace*{-2.5ex}
\begin{center}
  \begin{minipage}{6.5cm}
  \epsfig{figure=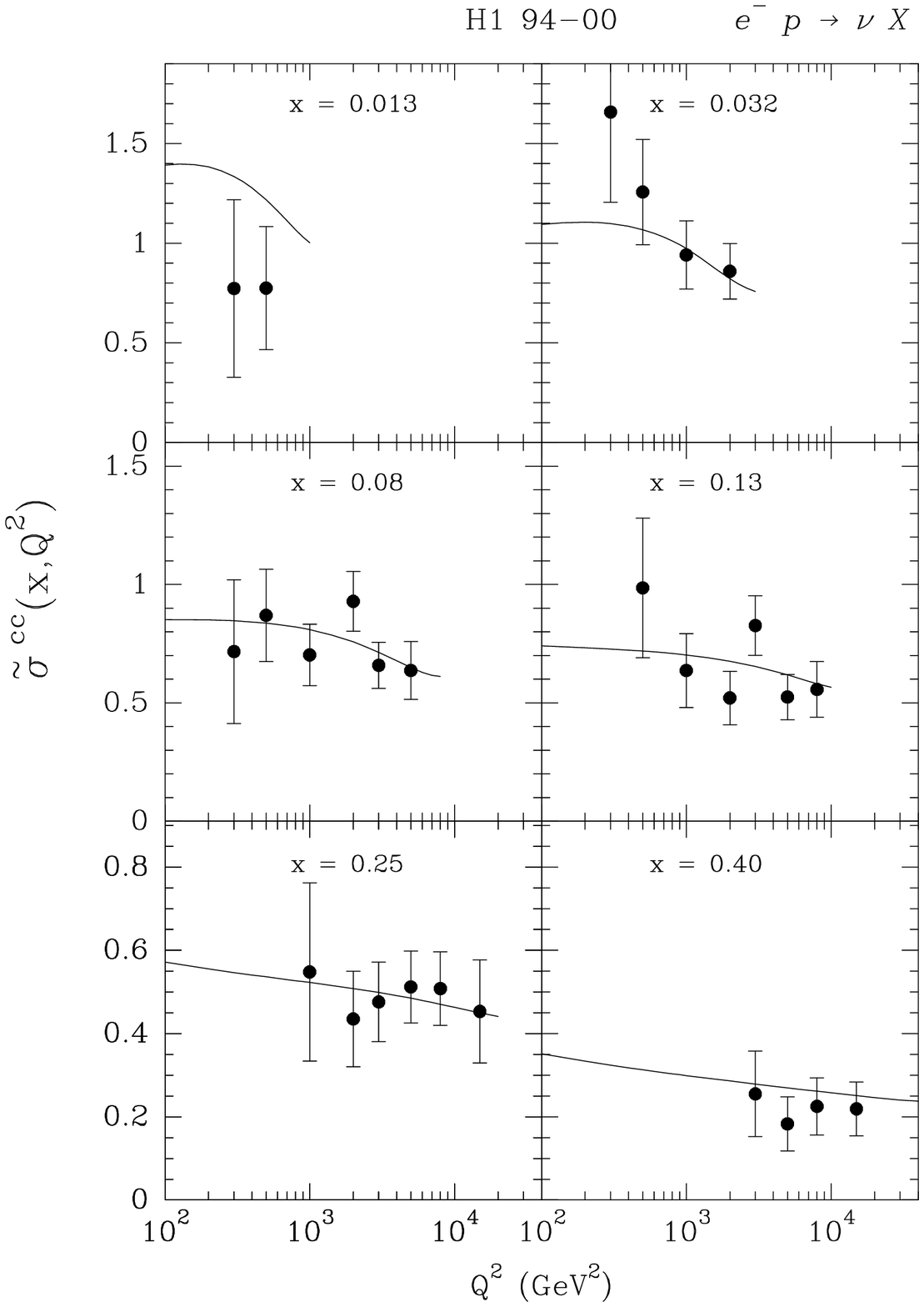,width=6.5cm}
  \end{minipage}
    \begin{minipage}{6.5cm}
  \epsfig{figure=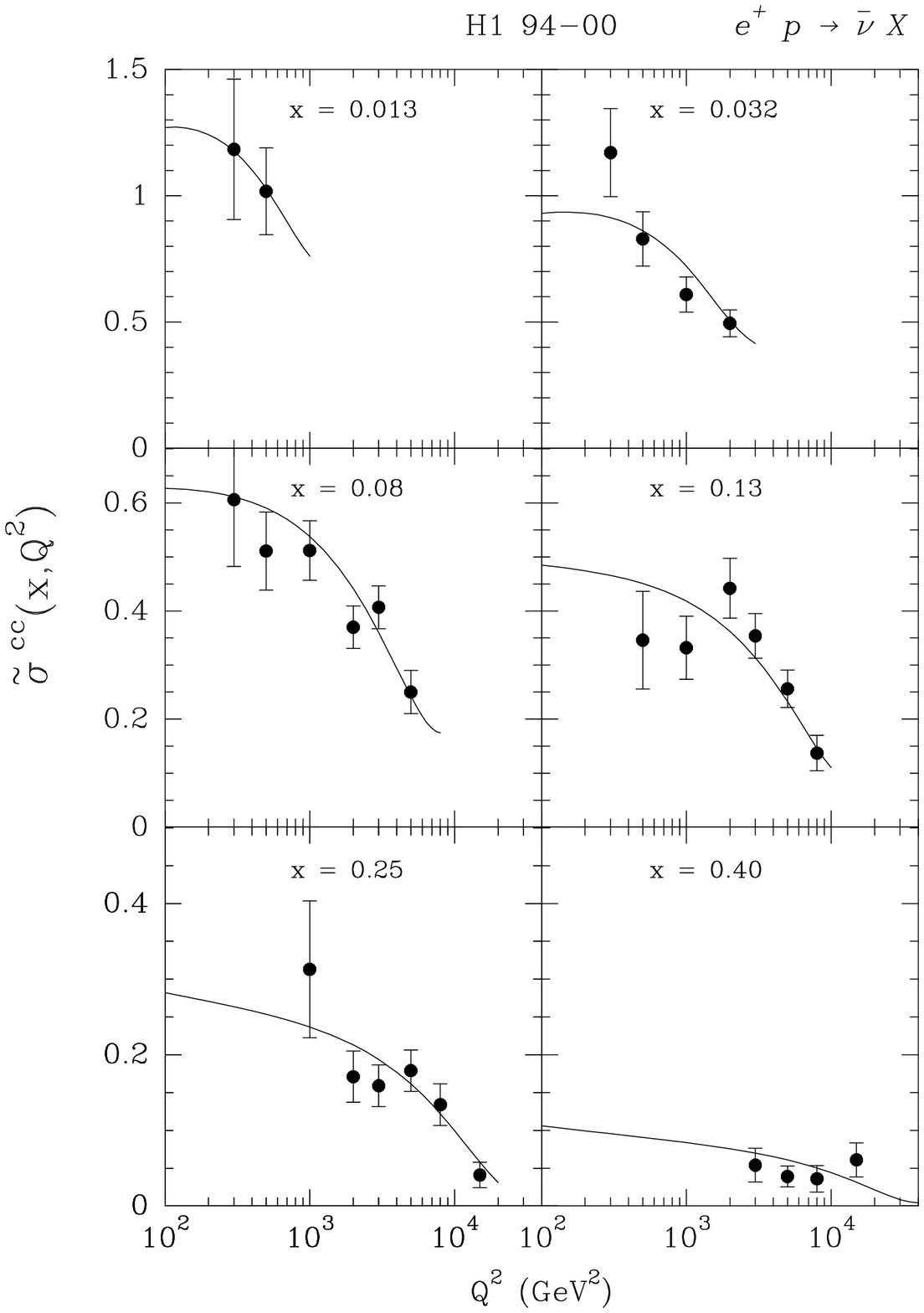,width=6.5cm}
    \end{minipage}
\end{center}
  \vspace*{-10mm}
\caption{
The reduced charged current cross section $\tilde{\sigma}$, 
in $e^{\pm} p$ reactions as a function of $Q^2$,
for different fixed values of $x$. Data from  H1 \cite{h103,h100}.}
\label{fi:he-p-vsq}
\vspace*{-2.5ex}
\end{figure}

\begin{figure}[t]
\begin{center}
  \begin{minipage}{6.5cm}
  \epsfig{figure=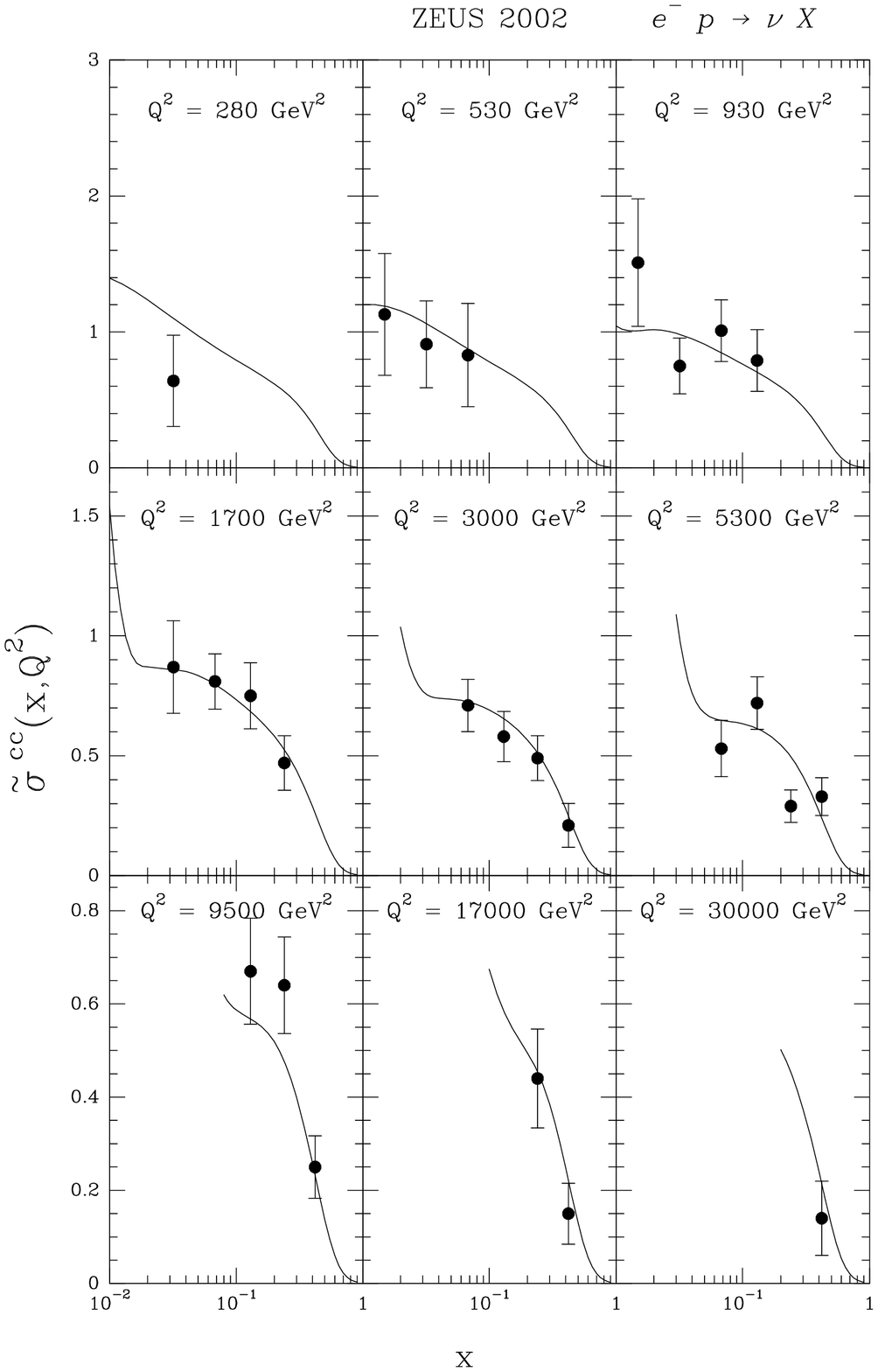,width=6.5cm}
  \end{minipage}
    \begin{minipage}{6.5cm}
  \epsfig{figure=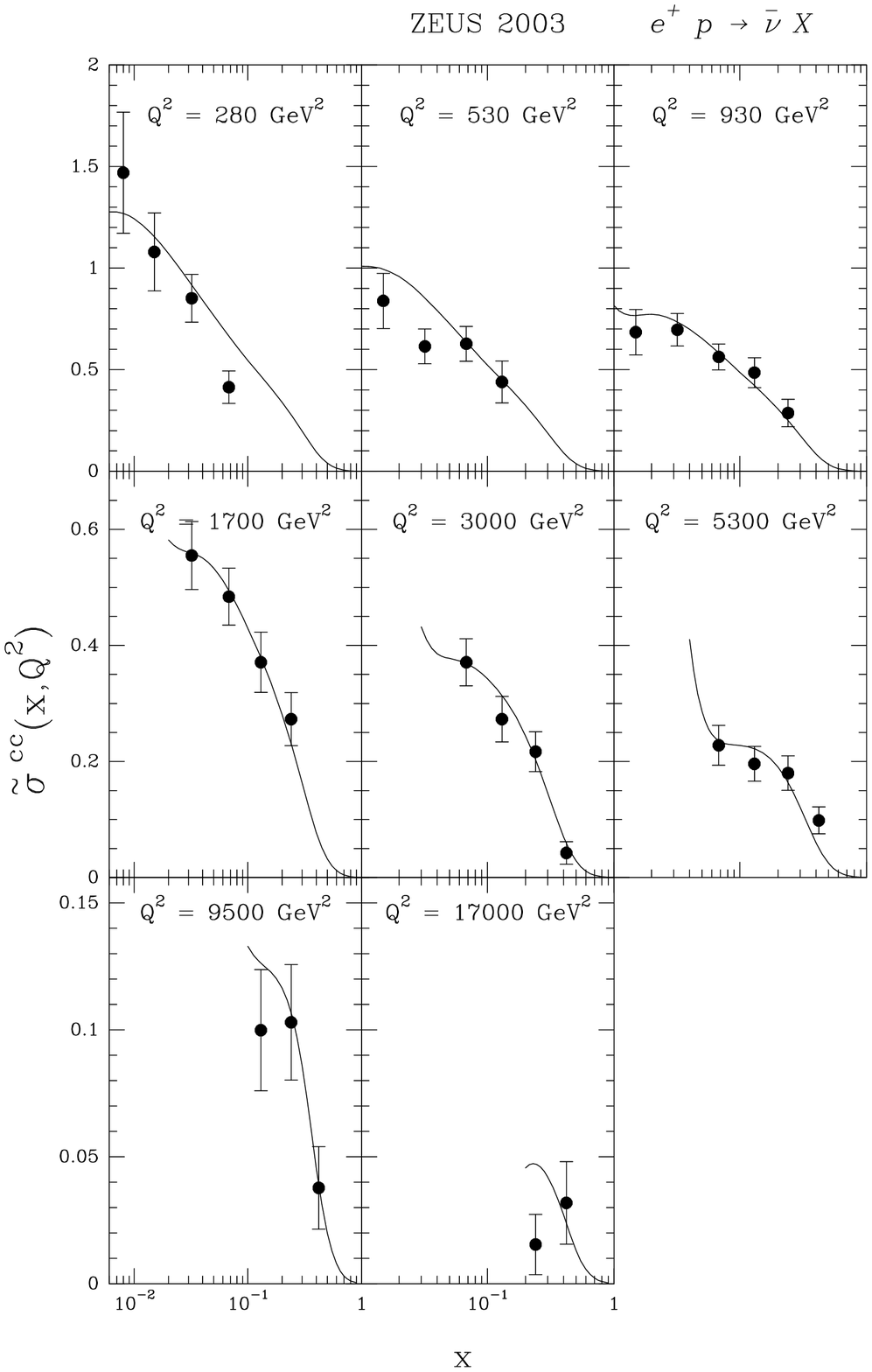,width=6.5cm}
    \end{minipage}\\
\end{center}
  \vspace*{-10mm}
\caption{The reduced charged current cross section, $\tilde{\sigma}$,
in $e^{\pm} p$ reactions as a function of $x$,
for different fixed values of $Q^2$. Data from ZEUS \cite{zeus03,zeus02a}.}
\label{fi:ze-p-vsx}
\vspace*{-2.5ex}
\begin{center}
  \begin{minipage}{6.5cm}
  \epsfig{figure=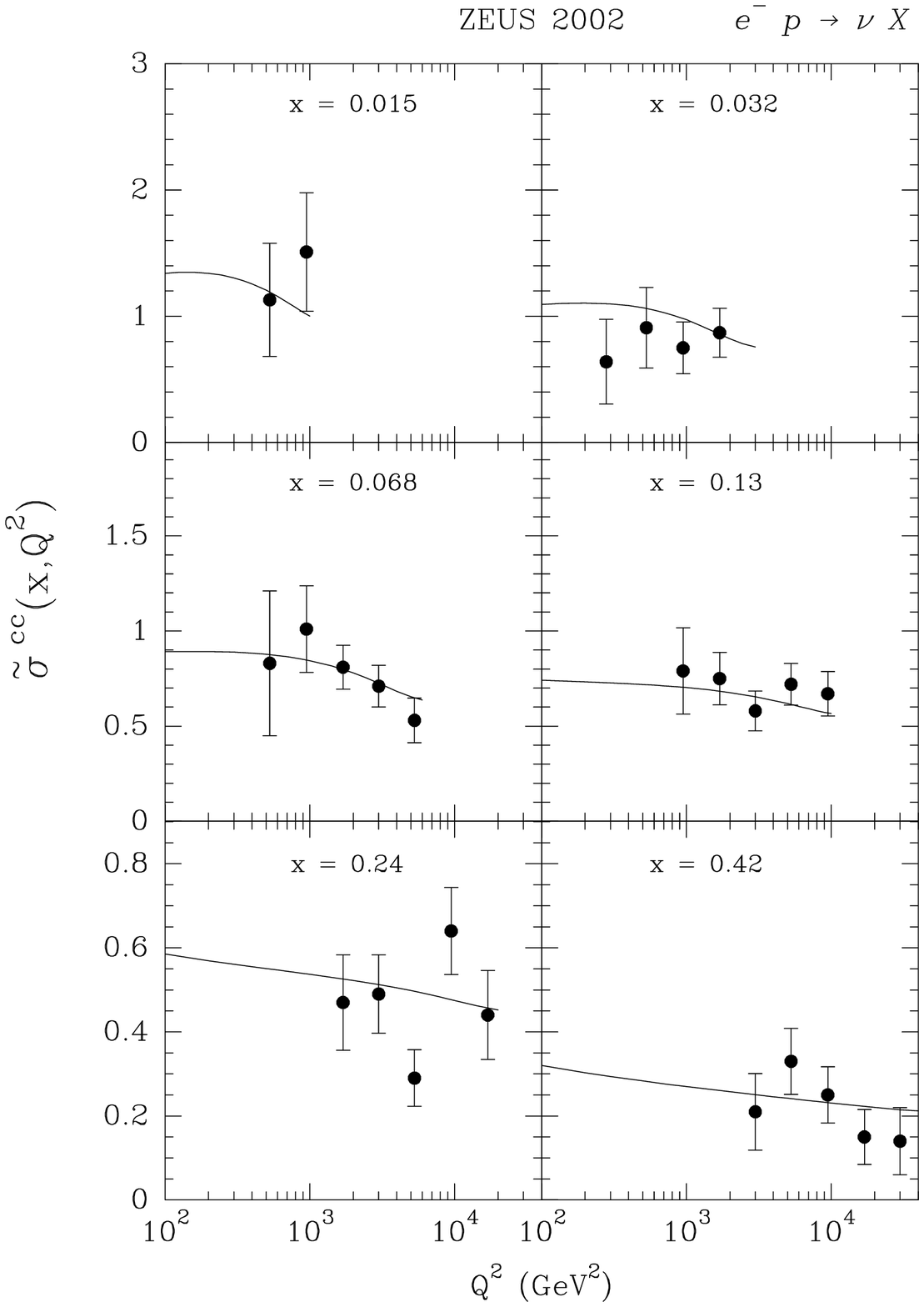,width=6.5cm}
  \end{minipage}
    \begin{minipage}{6.5cm}
  \epsfig{figure=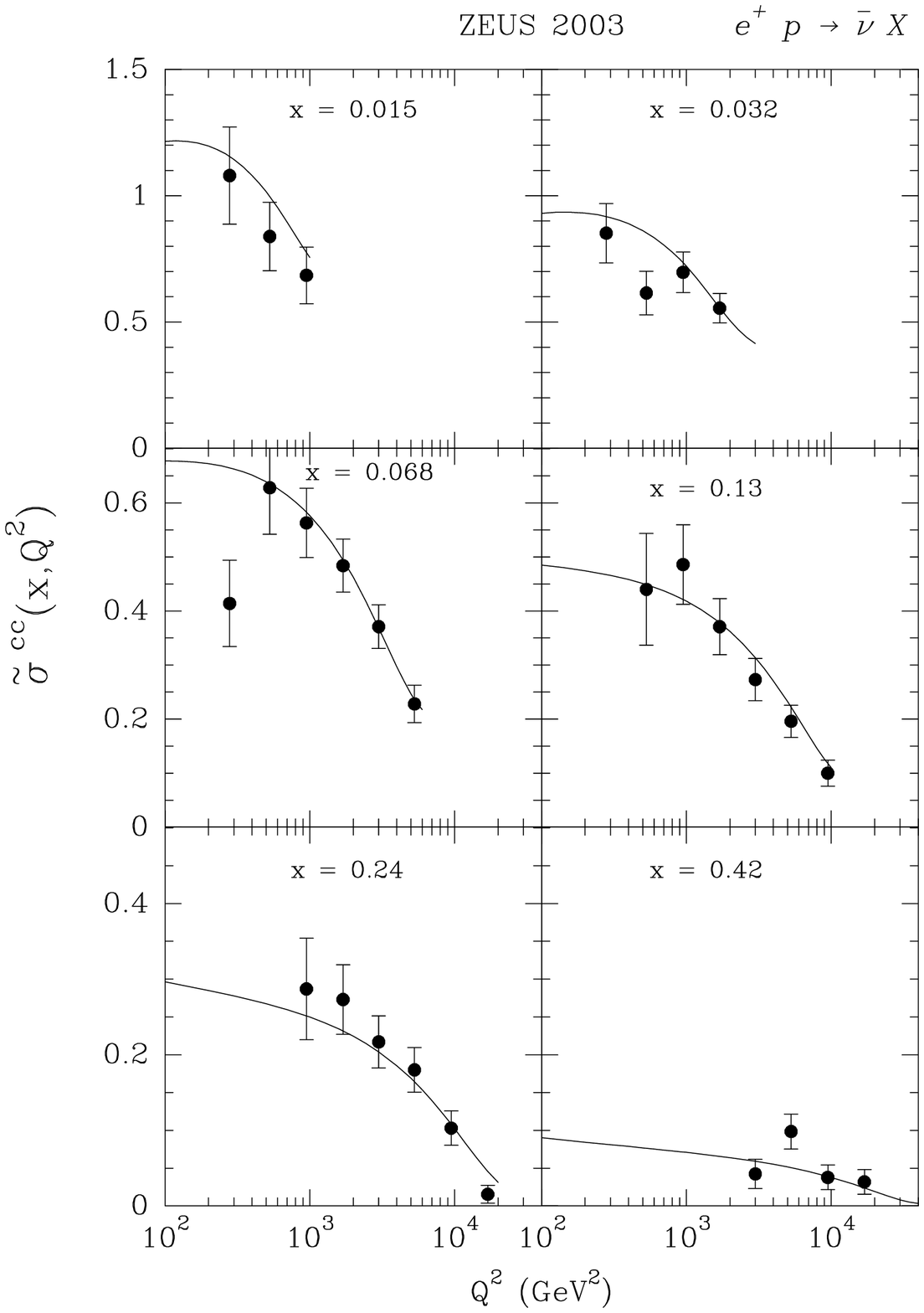,width=6.5cm}
    \end{minipage}
\end{center}
  \vspace*{-10mm}
\caption{
The reduced charged current cross section $\tilde{\sigma}$, 
in $e^{\pm} p$ reactions as a function of $Q^2$,
for different fixed values of $x$. Data from ZEUS \cite{zeus03,zeus02a}.}
\label{fi:ze-p-vsq}
\vspace*{-2.5ex}
\end{figure}

The charged current DIS processes have been also measured accurately at HERA
in an extented kinematic region. It has a serious impact on the determination
of the unpolarized parton distributions by allowing a flavor separation because
they involve only the $W^{\pm}$ exchange. The cross sections are expressed, at
lowest order, in terms of three structure functions as follows \cite{zhang}
\begin{eqnarray}
\frac{d^2 \sigma^{cc}_{Born}}{dx dQ^2} &=&
\frac{G_F^2}{4\pi}\frac{M^4_W}{(Q^2 + M^2_W)^2}
\left[Y_+ F^{cc}_2 (x, Q^2) - y^2 F^{cc}_L(x, Q^2) \right. \nonumber \\
&&\left. + Y_- xF^{cc}_3(x, Q^2) \right]~,
\label{chargcur}
\end{eqnarray}
and the reduced cross sections are defined as
\begin{equation}
\tilde \sigma^{cc}(x, Q^2) = 
\left[\frac{G_F^2}{4\pi}\frac{M^4_W}{(Q^2 + M^2_W)^2} \right]^{-1}
\frac{d^2 \sigma^{cc}}{dx dQ^2}~.
\label{redcross}
\end{equation}
At leading order for $e^- p \rightarrow \nu_e X$ with a longitudinally 
polarized beam
\begin{eqnarray}
F^{cc}_2(x, Q^2) &=& x[u(x, Q^2) + c(x, Q^2) + \bar d(x, Q^2) +
\bar s (x, Q^2)] \nonumber \\
xF^{cc}_3(x, Q^2) &=& x[u(x, Q^2) + c(x, Q^2) - \bar d(x, Q^2)
-\bar s(x, Q^2)]~,
\label{f2em}
\end{eqnarray}
and for $e^+ p \rightarrow \bar \nu_e X$
\begin{eqnarray}
F^{cc}_2(x, Q^2) &=& x[d(x, Q^2) + s(x, Q^2) + \bar u(x, Q^2) +
\bar c (x, Q^2)] \nonumber \\
xF^{cc}_3(x, Q^2) &=& x[d(x, Q^2) + s(x, Q^2) - \bar u(x, Q^2)
-\bar c(x, Q^2)]~.
\label{f2ep}
\end{eqnarray}
At NLO in QCD $F^{cc}_L$ is non zero, but it gives negligible contribution, 
except at $y$ values close to 1.
Our predictions are compared with H1 and ZEUS data in Figs.~\ref{fi:he-p-vsx},
\ref{fi:he-p-vsq}, \ref{fi:ze-p-vsx} and \ref{fi:ze-p-vsq}, as a function of 
$x$ in a broad range of $Q^2$ values and vice versa. 
The agreement is very good, but unfortunately since the highest $x$ value is
only 0.42, it does not allow to clarify the situation regarding the large 
$x$ behavior, as already noticed above.

\clearpage
\newpage

\newpage

\newpage
\newpage
\clearpage

\section{Charged current neutrino cross sections}
\begin{figure}[t]
\vspace*{-5ex}
\begin{center}
  \begin{minipage}{6.5cm}
  \epsfig{figure=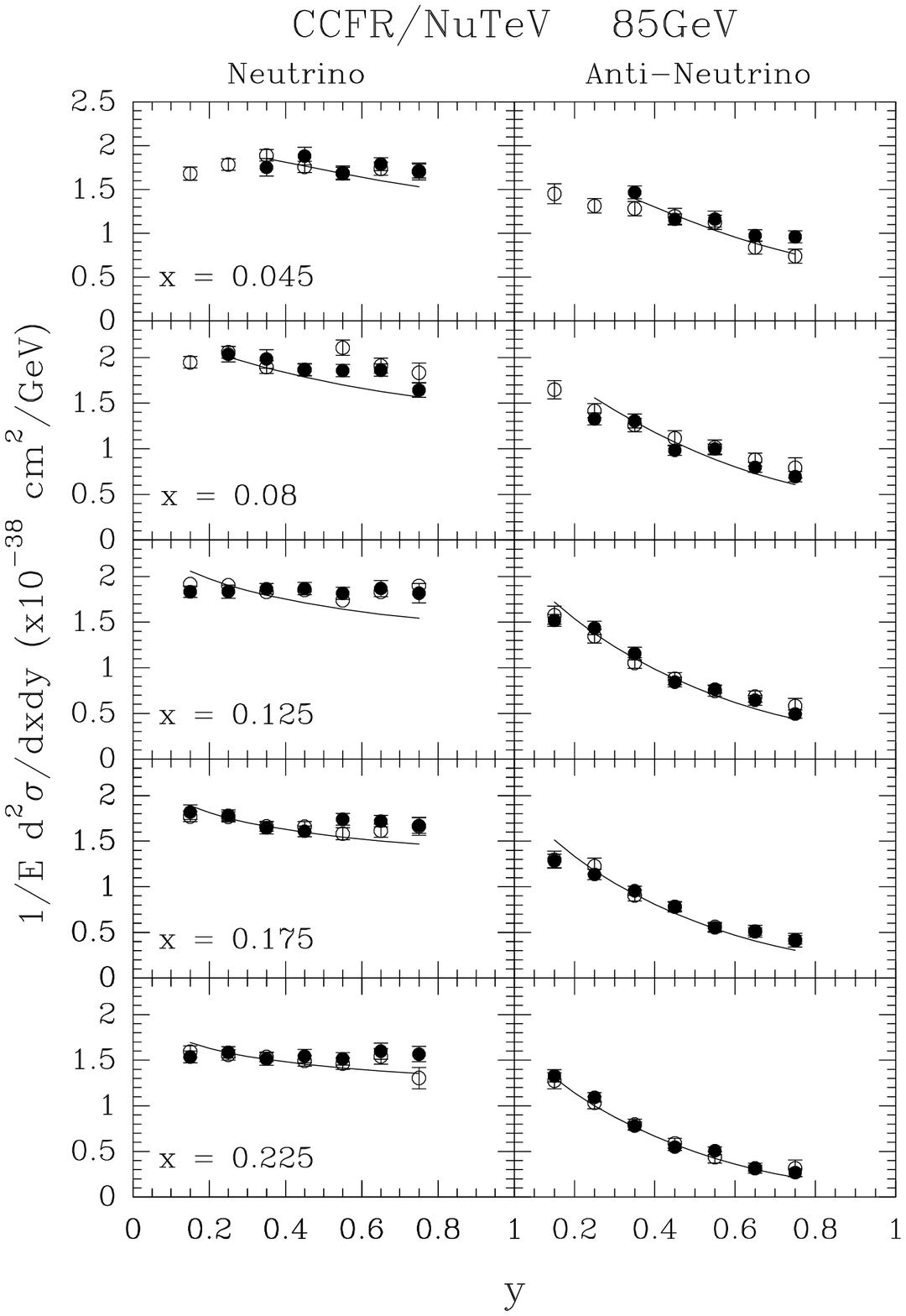,width=7.5cm}
  \end{minipage}
    \begin{minipage}{6.5cm}
  \epsfig{figure=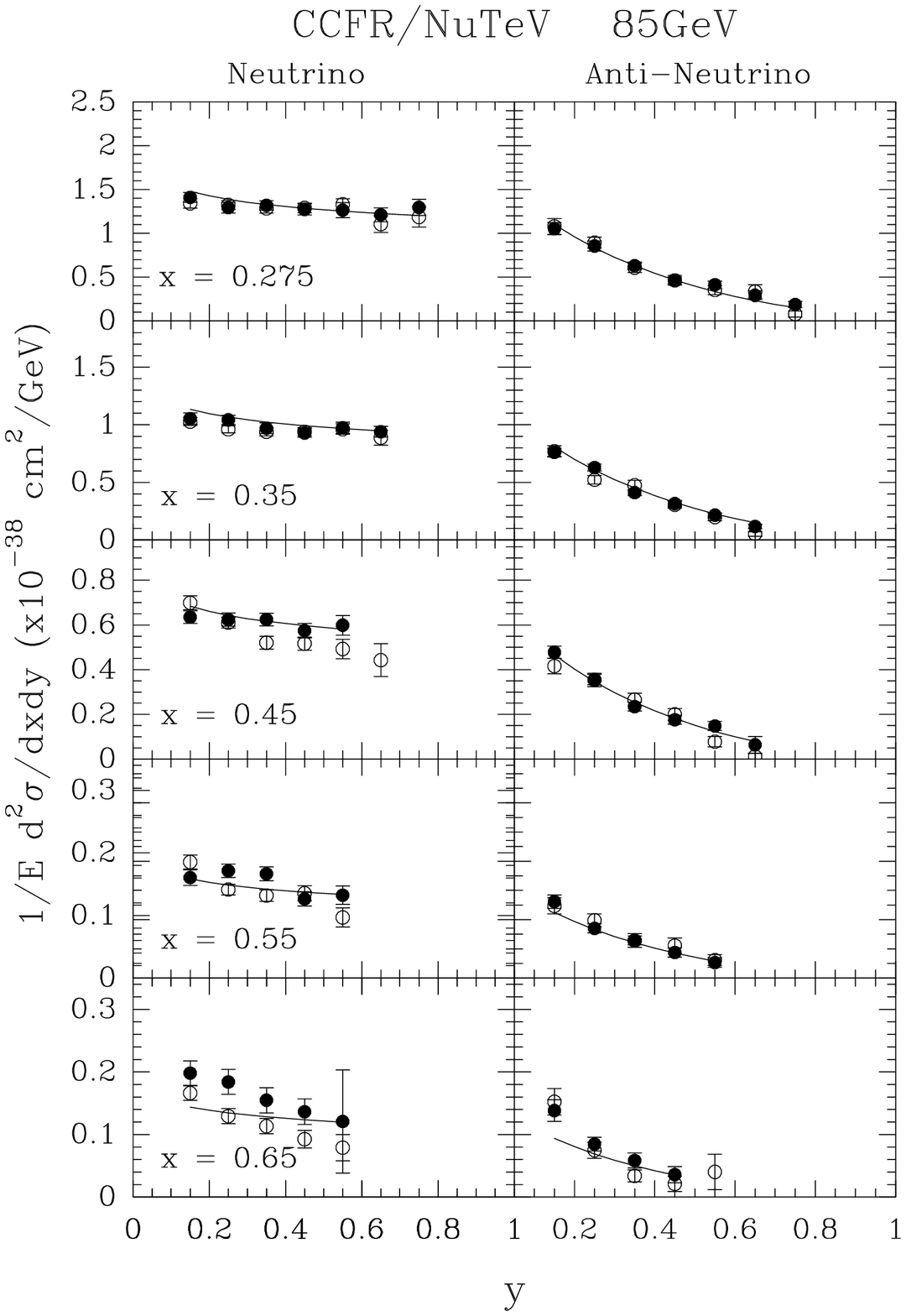,width=7.5cm}
    \end{minipage}
\end{center}
  \vspace*{-15mm}
\caption{
Differential cross section $\nu (\bar \nu )N$ for 
$E_{\nu} = 85 \mbox{GeV}$, as
a function of $y$. Data are from CCFR \cite{UKYang} (white circles)
and NuTeV experiments \cite{nutevtalk,nutevdat} (black circles).}
\label{fi:enu85}
\vspace*{-4.5ex}
\begin{center}
\begin{minipage}{6.5cm}
 \epsfig{figure=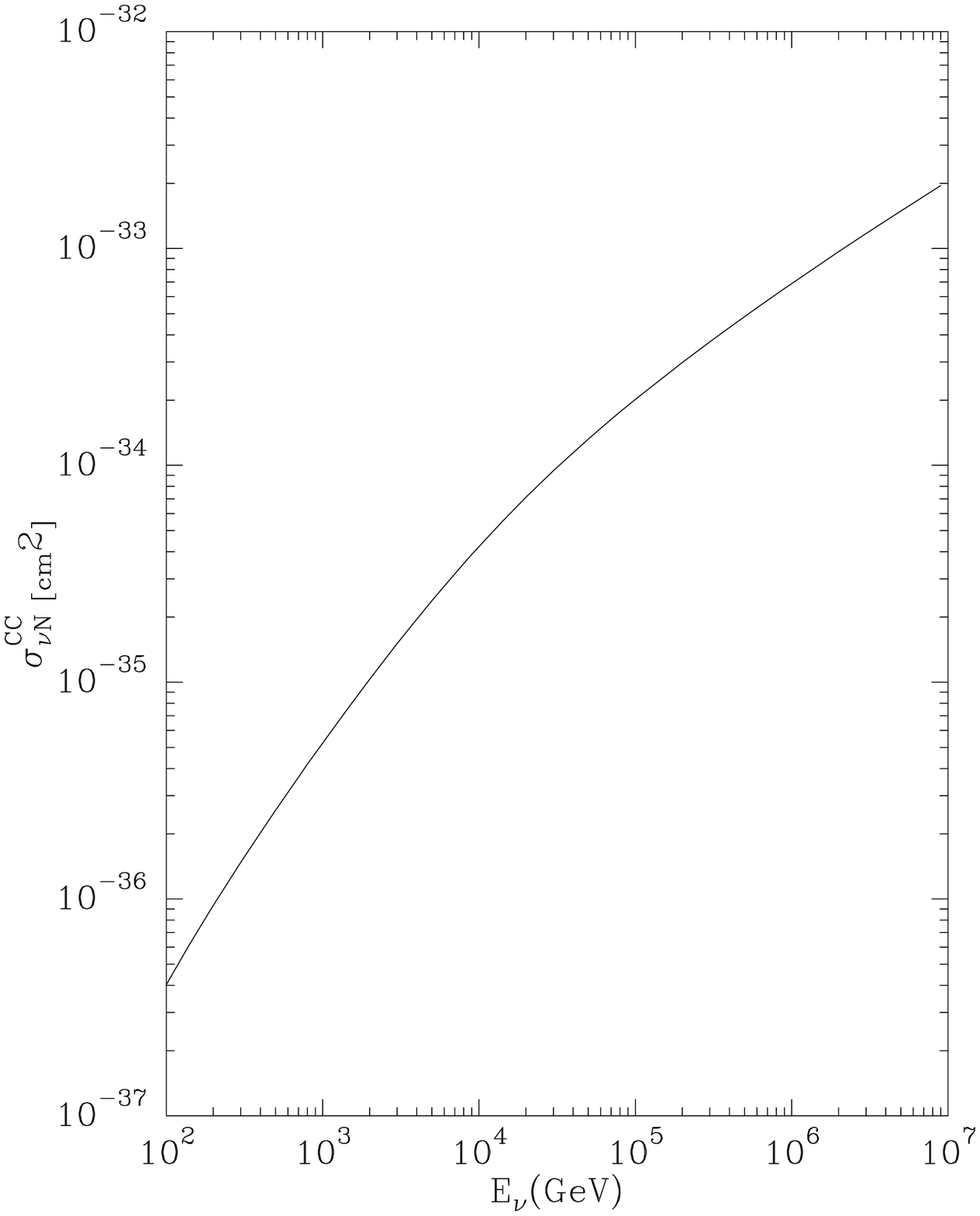,width=7.5cm}
 \end{minipage}
\end{center}
  \vspace*{-20mm}
\caption[*]{\baselineskip 1pt
Charged current total cross section $\nu N$, 
for an isoscalar target as a function of the neutrino energy.}
\label{fi:nun}
\vspace*{-2.5ex}
\end{figure}

The differential inclusive neutrino and antineutrino cross sections have the 
following standard expressions
\begin{eqnarray}
\frac{d^2 \sigma^{\nu, (\bar \nu)}}{dx dy} &=&
\frac{G^2_F M_p E_{\nu}}{ \pi (1 + \frac{Q^2}{M^2_W})^2}
\left[x y^2 F_1^{\nu (\bar \nu)}(x, Q^2)
+ (1-y-\frac{M_p xy}{2E_{\nu}})
F_2^{\nu (\bar \nu)}(x, Q^2) \right. \nonumber \\
&& \left. \pm (y - \frac{y^2}{2})x F_3^{\nu (\bar \nu)}(x, Q^2)\right]\,,
\label{dcrossnu}
\end{eqnarray}
$y$ is the fraction of total leptonic energy transfered to the 
hadronic system and $E_{\nu}$ is the incident neutrino  energy.
$F_2$ and $F_3$ are given by
Eq.~(\ref{f2em}) for $\nu p$ and Eq.~(\ref{f2ep}) for $\bar \nu p$, and $F_1$
is related to $F_2$ by
\begin{equation}
2 x F_1 = \frac{1 + 4 M^2_p x^2}{1 + R} F_2\,,
\label{f1vf2}
\end{equation}
where $R = \sigma_L / \sigma_T$, the ratio of the longitudinal to
transverse cross sections of the W-boson production. The calculations
are done with $\sin^2{\theta_W} = 0.2277\pm 0.0013 \pm 0.009$ 
obtained by NuTeV \cite{zeller} and the comparison with the CCFR and NuTeV
data are shown in Fig.~\ref{fi:enu85}. As expected, for fixed $x$, the $y$ 
dependence is rather flat for neutrino and has the characteristic 
$(1-y)^2$ behavior for antineutrino.\\ 

This can be extrapolated to evaluate the cross section of ultrahigh energy 
neutrinos with nucleons. The total cross section at a given neutrino energy 
reads
\begin{equation}
\sigma^{CC}_{\nu N}(E_{\nu}) = \int dx dy 
\frac{d^2 \sigma^{\nu, (\bar \nu)}}{dx dy}~.
\label{crossnu}
\end{equation}
Our prediction for this total charged current cross section, for an isoscalar
nucleon $ N=1/2(p+n)$, versus the neutrino energy, is displayed in 
Fig.~\ref{fi:nun} and it has the expected strong energy increase.
We have not calculated the corresponding neutral current cross section, which
is known to be a factor three or so smaller. This new information is certainly
valuable to the large scale neutrino telescopes, for the detection of
extraterrestrial neutrino sources.

\clearpage
\newpage

\section{Drell-Yan dilepton cross sections}
\begin{figure}
\begin{center}
\begin{minipage}{6.5cm}
\epsfig{figure=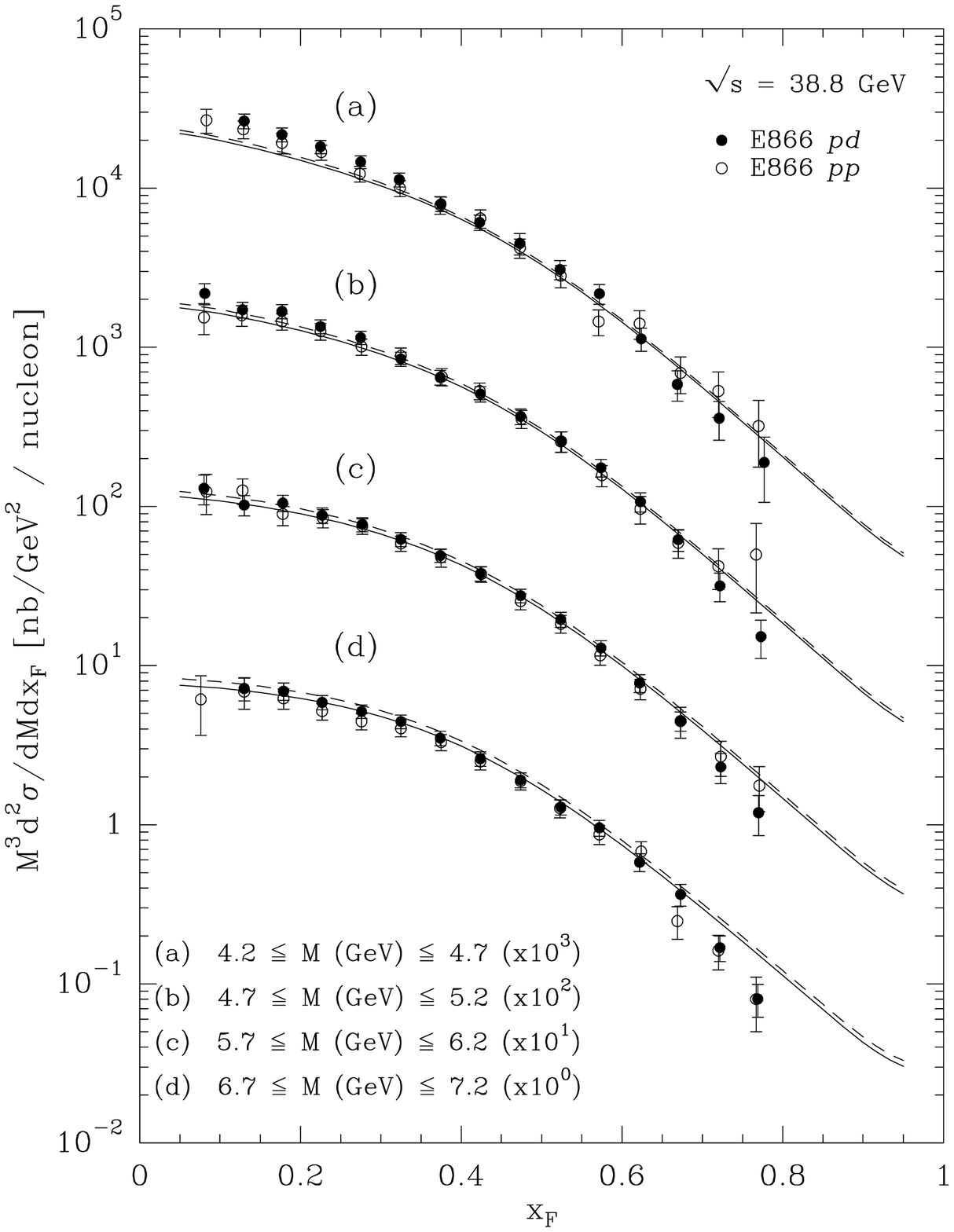,width=8.0cm}
\end{minipage}
\begin{minipage}{6.5cm}
\epsfig{figure=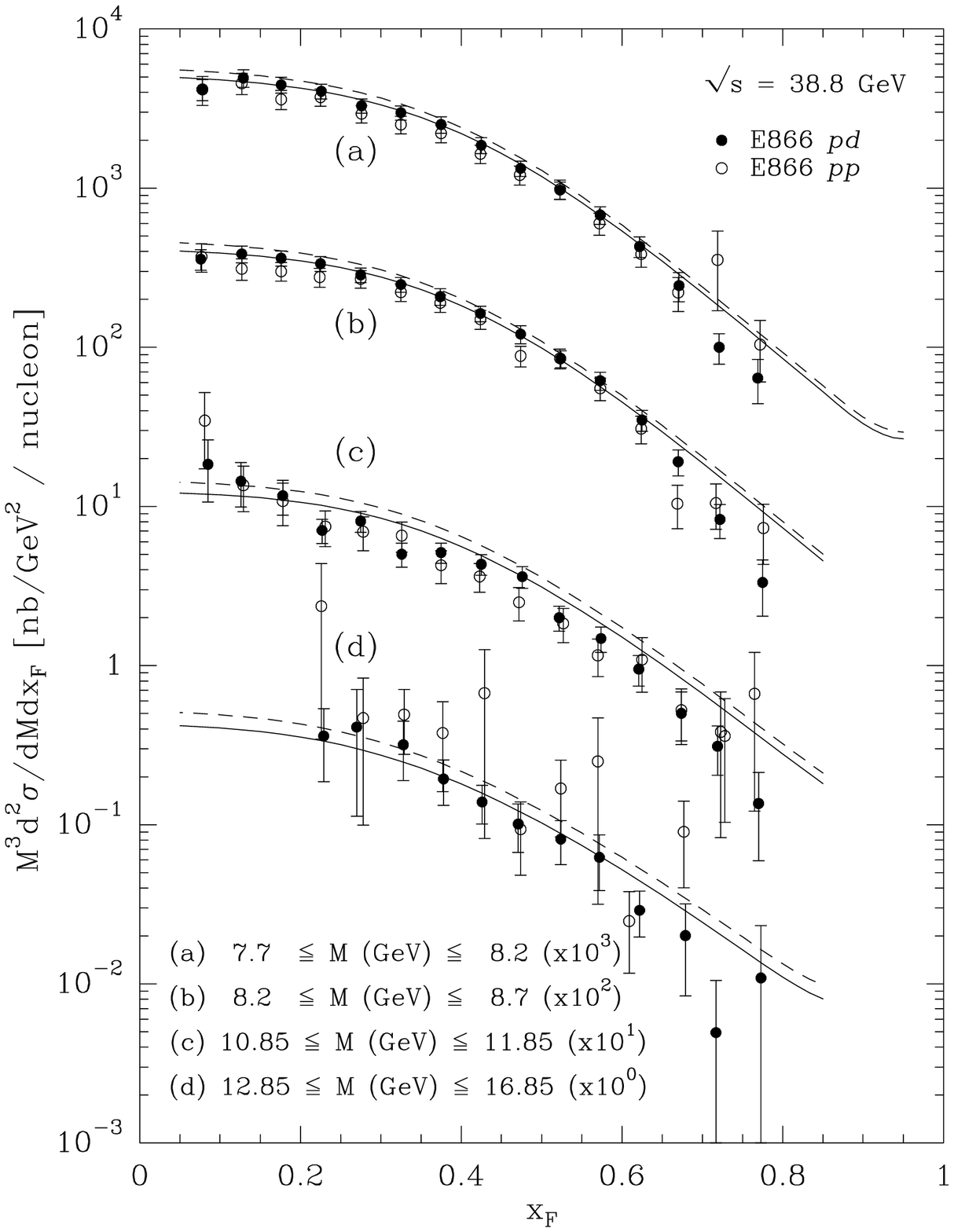,width=8.0cm}
\end{minipage}
\end{center}
\vspace*{-15mm}
\caption{
Drell-Yan cross sections per nucleon at $\sqrt{s} = 38.8\mbox{GeV}$
for $p p$ and  $p d$  as a function of $x_F$ for selected $M$ bins.
Solid curve $pp$, dashed curve $pd$.
Experimental data from FNAL E866 \cite{E866a}.}
\label{fi:drell1xf}
\vspace*{-1.5ex}
\end{figure}

\begin{figure}
\begin{center}
\epsfig{figure= 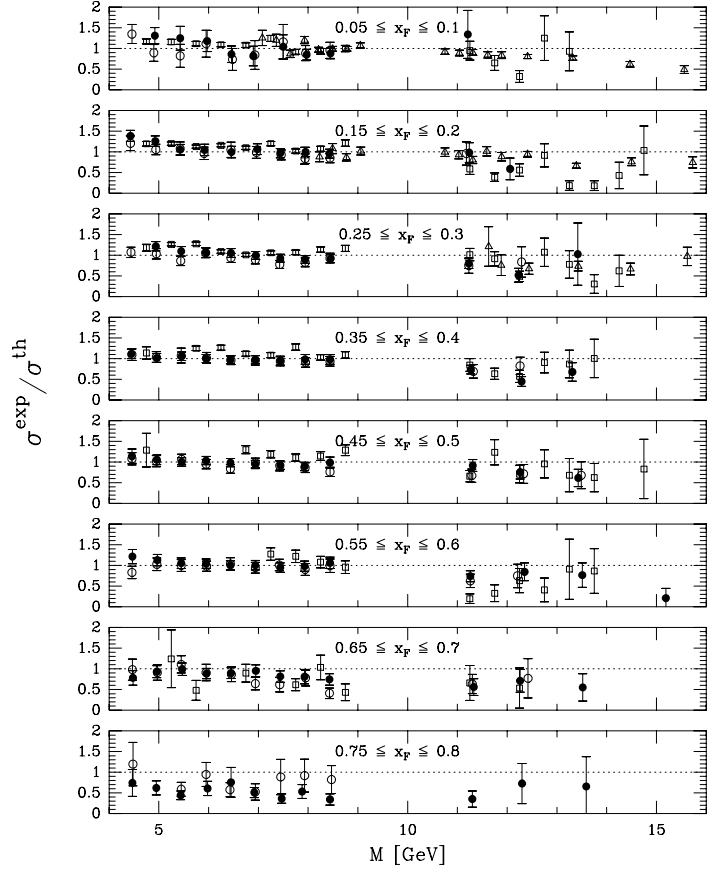,width=11.5cm}
\end{center}
\vspace*{-15mm}
\caption[*]{\baselineskip 1pt
Drell-Yan cross sections ratios experiment versus theory
at $\sqrt{s} = 38.8\mbox{GeV}$
for $p p$ (open circle), $p d$ (full circle, square), and $p Cu$ (triangle)
as a function of $M$ for selected $x_F$ bins.
Experimental data are from Refs. \cite{E866a,E605,E772}.}
\label{fi:drellratio}
\vspace*{-1.5ex}
\end{figure}

A very important source of information for $\bar{q}(x)$ distributions
comes from Drell-Yan dilepton processes, whose cross sections are proportional
to a combination of products of $q(x)$ and $\bar{q}(x)$ distributions. The 
cross section $\sigma_{DY}(pp)$ for $pp \to \mu^+\mu^- X$, at the lowest
order, has the simplified form
\begin{equation}
M^3\frac{d^2 \sigma_{DY}(pp)}{dM dx_F} = \frac{8\pi \alpha^2}{9(x_1+x_2)}
\sum_i e_i^2 [q_i(x_1)\bar q_i(x_2) + \bar q_i(x_1) q_i(x_2)]~,
\label{crossdy1}
\end{equation}
where $M$ is the invariant mass of the produced muon pair, $x_1$ and $x_2$ 
refer to the beam and target respectively, 
$x_F=x_1-x_2$ and $M^2=x_1 x_2 s$,  where $\sqrt{s}$ is the center of mass 
energy of the collision. Clearly at NLO one should add the Compton processes 
contributions to the above $q \bar{q}$ annihilation terms.\\
More recently the NuSea Collaboration has released the data on the Drell-Yan 
cross sections $\sigma_{DY}(pp)$ and $\sigma_{DY}(pd)$ for proton-proton 
and proton-deuterium collisions at 800 GeV/c \cite{E866a}. 
They are displayed in Fig.~\ref{fi:drell1xf} as a function of $x_F$ for 
selected $M$ bins, together with our predictions. The agreement is
fairly good, mainly in the small mass region, but in order to evaluate it 
more precisely, we have plot in Fig.~\ref{fi:drellratio}
the ratios of experiment versus theory, using a broader set of data.\\
Let us now come back to the extraction of the ratio $\bar d/\bar u$ from these
data. For large $x_F$, namely $x_1>>x_2$ and small $M$, we have
\begin{equation}
\frac{\sigma_{DY}(pd)}{ 2 \sigma_{DY}(pp)} \simeq 1/2 \left[ 1 + 
\frac{\bar d(x_2)}{\bar u(x_2)}\right]~,
\label{pdpp1}
\end{equation}
so the measurement of this cross sections ratio is directly related to 
$\bar d(x)/\bar u(x)$ for small $x$. For large $x$ one needs to use small
$x_F$ and large $M$ values and we have now for $x_1\simeq x_2$
\begin{equation}
\frac{\sigma_{DY}(pd)}{ 2 \sigma_{DY}(pp)} \simeq 1/2 \left[ \frac{8 + 
5 \frac{\bar d(x)}{\bar u(x)} +
5\frac{d(x)}{u(x)} + 2 \frac{\bar d(x)}{\bar u(x)} \frac{d(x)}{u(x)}}{8 + 
2 \frac{\bar d(x)}{\bar u(x)} \frac{d(x)}{u(x)}}\right]~.
\label{pdpp2}
\end{equation}
Therefore the falloff at large $x$ of $\sigma_{DY}(pd)/2\sigma_{DY}(pp)$
observed in Ref.\cite{E866} cannot be directly related to the falloff 
of $\bar d(x)/\bar u(x)$, since $d(x)/u(x)$ is also decreasing for large $x$, 
as shown previously (see Fig.~\ref{fi:doveru}). The use of Eq.~(\ref{pdpp1})
will lead to an underestimation of $\bar d(x)/\bar u(x)$. We also notice in
Fig.~\ref{fi:drell1xf} an experimental point for $\sigma_{DY}(pp)$ in 
the bin with $M$ in the range (10.85,11.85)GeV at $x_F\simeq 0.05$, two 
standard deviations above our curve, which might very well
be one of the reason for the dramatic falling off of $\bar d(x)/\bar u(x)$ 
for $x\simeq 0.3$, reported by NuSea. Obviously more accurate data in 
this region are badly needed.

\clearpage
\newpage

\section{Single jet and $\pi^0$ inclusive productions}
A precise determination of parton distributions allows us to use them
as input information to predict strong interaction processes, for 
additional tests of pertubative QCD and also for the search of
new physics. Here we shall test our statistical parton distributions for the  
description of two inclusive reactions, single jet and $\pi^0$ productions.
The cross section for the production of a single jet of rapidity $y$ and 
transverse momentum $p_T$, in a $\bar{p}p$ collision is given by  
\begin{eqnarray}
E\frac{d^3\sigma}{dp^3} 
&=& \sum_{ij}\frac{1}{1+\delta_{ij}}\frac{2}{\pi}
\int_{x_0}^{1} dx_a \frac{x_a x_b}{2x_a - x_Te^y} \times
\nonumber\\
&&\left[f_i(x_a, Q^2) f_j(x_b, Q^2)\frac{d\hat \sigma_{ij}}{d\hat{t}}(\hat s,
\hat t, \hat u) + (i\leftrightarrow j) \right]~,
\label{crossjet}
\end{eqnarray}
where $x_T = 2p_T / \sqrt{s}$, $x_0 = x_T e^y/(2 - x_T e^{-y})$,
$x_b = x_a x_T e^{-y}/(2x_a - x_T e^y)$ and $\sqrt{s}$
is the center of mass energy of the collision. In the above sum, $i,j$ stand 
for initial gluon-gluon, quark-gluon and quark-quark scatterings, 
$d\hat \sigma_{ij}/d\hat{t}$ are the corresponding partonic cross sections 
and $Q^2$ is the scaling variable.
The NLO QCD calculations at ${\cal O} ( \alpha_s^3 )$, were done using a code
described in Ref.~\cite{JSV}, based on a semi-analytical method within the
"small-cone approximation"\footnote{ We thank Werner Vogelsang for providing 
us with the numerical values, resulting from the use of our 
parton distributions}.

\begin{figure}[htb]
\vspace*{-20mm}
\begin{center}
\leavevmode {\epsfysize= 14.0cm \epsffile{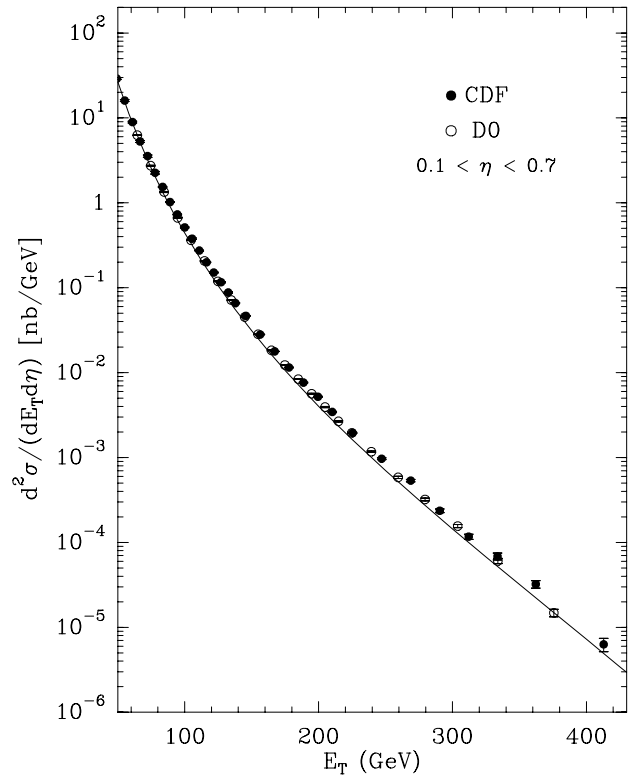}}
\end{center}
\vspace*{-20mm}
\caption[*]{\baselineskip 1pt
Cross section for single jet production in $\bar p p$ at $\sqrt{s} = 1.8
\mbox{TeV}$ as a function of $E_T$. Data are from CDF \cite{CDF2001}
and D0 \cite{D02001} experiments.}
\label{fi:jet}
\vspace*{-2.0ex}
\end{figure}

\begin{figure}
\vspace*{-20mm}
\begin{center}
\leavevmode {\epsfysize= 14.0cm \epsffile{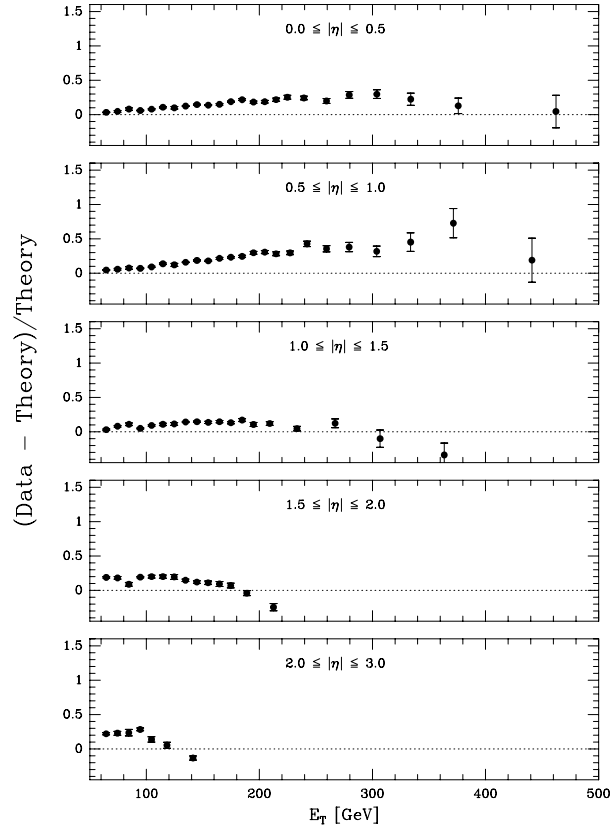}}
\end{center}
\vspace*{-20mm}
\caption[*]{\baselineskip 1pt
Comparison between the statistical model and the D0 \cite{D02001a}
single jet cross sections in $\bar p p$ at $\sqrt{s} = 1.8
\mbox{TeV}$, as a function of $E_T$ and rapidity $\eta$.}
\label{fi:rapjet}
\vspace*{-2.0ex}
\end{figure}

In Fig.~\ref{fi:jet} our results are compared with the data from CDF and D0
experiments \cite{CDF2001,D02001}. Our prediction agrees very well with the 
data up to the highest $E_T$ (or $p_T$) value and this is remarkable given 
the fact that the experimental results are falling off over more than 
six orders of magnitude, leaving no room for new physics. For completeness,
we also show in Fig.~\ref{fi:rapjet} the D0 data, for several rapidity bins, 
using a (Data-Theory)/Theory presentation.\\

\begin{figure}
\begin{center}
\leavevmode {\epsfysize= 14.cm \epsffile{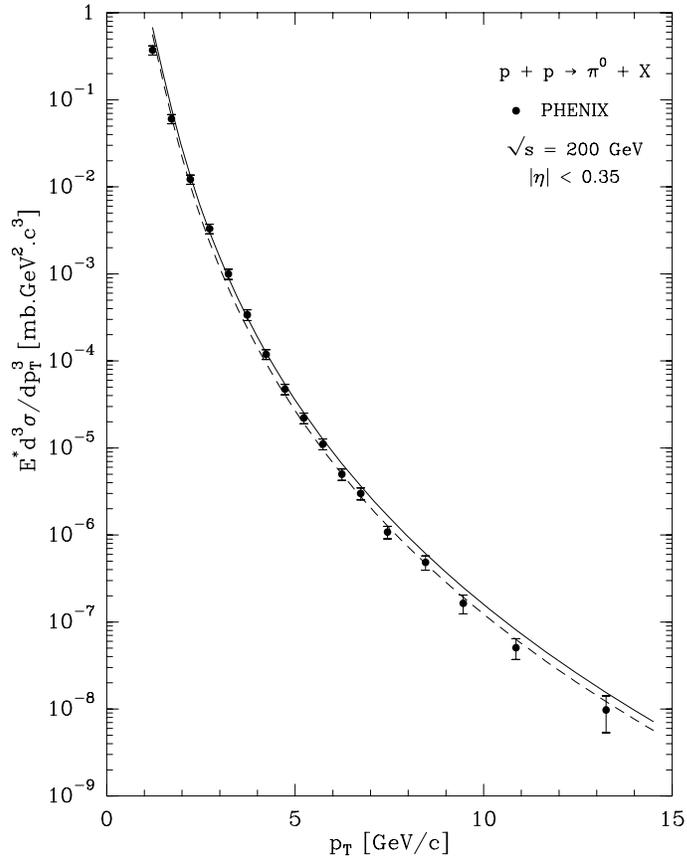}}
\end{center}
\vspace*{-5mm}
\caption[*]{\baselineskip 1pt
Inclusive $\pi^0$ production in $p p$ reaction at $\sqrt{s} = 200
\mbox{GeV}$ as a function of $p_T$, scale $\mu = p_T$. 
Data from PHENIX \cite{phenix03}. 
Solid curve fragmentation functions from KKP \cite{kniel00},
dashed curve from BKP \cite{binew95}.}
\label{fi:pi0phenix}
\vspace*{-1.5ex}
\end{figure}

\begin{figure}
\begin{center}
\leavevmode {\epsfysize=14.0cm \epsffile{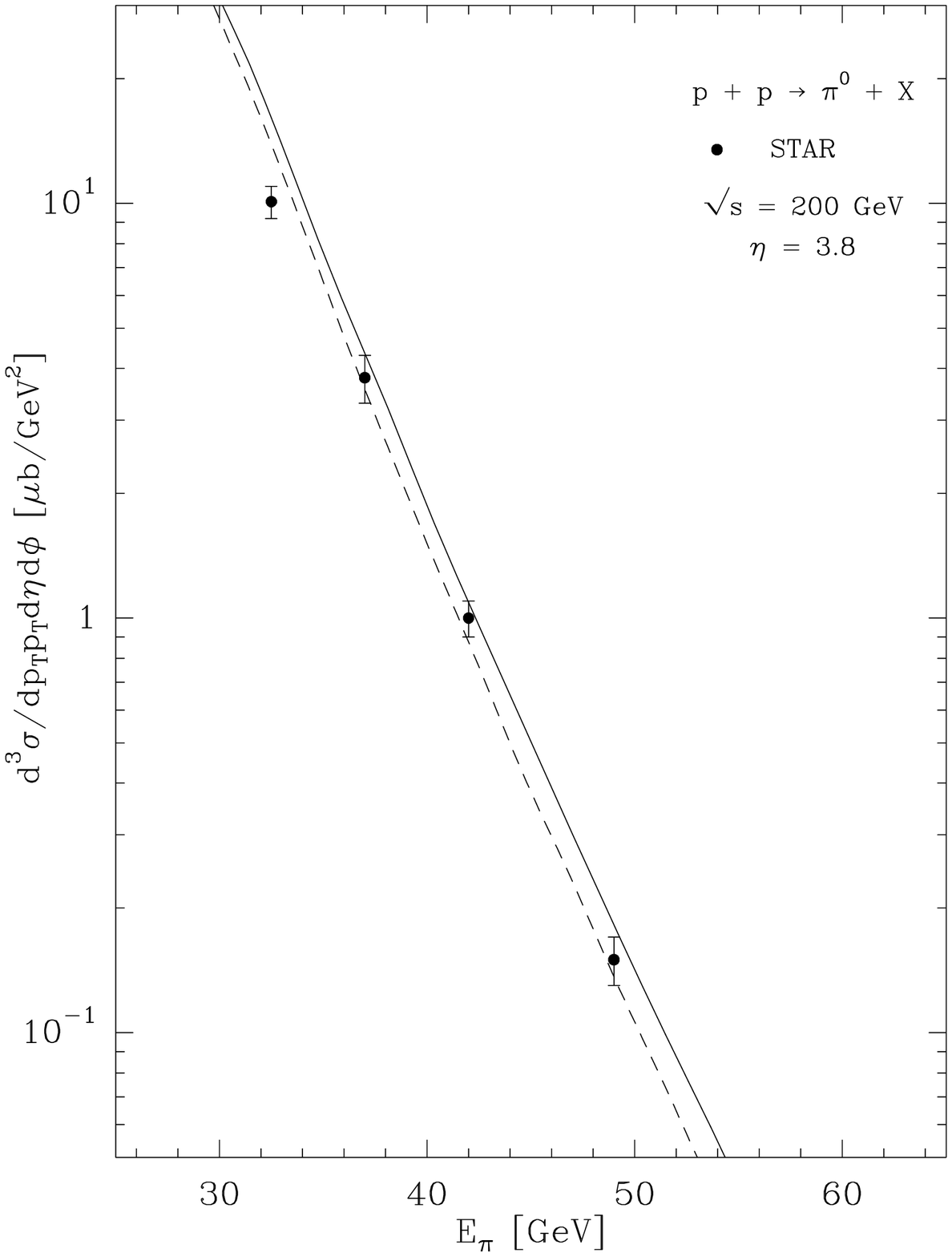}}
\end{center}
\vspace*{-5mm}
\caption[*]{\baselineskip 1pt
Inclusive $\pi^0$ production in $p p$ reaction at $\sqrt{s} = 200
\mbox{GeV}$ as a function of $E_{\pi}$. 
Data from STAR  \cite{star03}. 
Solid curve fragmentation functions from KKP \cite{kniel00},
dashed curve from BKP \cite{binew95}.}
\label{fi:pi0star}
\vspace*{-1.5ex}
\end{figure}

Next we consider the cross section for the inclusive production of a $\pi^0$ 
of rapidity $y$ and transverse momentum $p_T$, in a $pp$ collision, which
has the following expression
\begin{eqnarray}
E_{\pi} d^3\sigma / dp_{\pi}^3
&=& \sum \limits_{abc} \int dx_a dx_b\, f_{a/p}(x_a,Q^2) \times  
\nonumber \\ 
&&f_{b/p}(x_b,Q^2)\frac{ D_{\pi^0/c}(z_c,Q^2)} 
{\pi z_c}d\hat {\sigma} /d \hat{t}(ab \to cX)~,
\label{Dsig}
\end{eqnarray}
where the sum is over all the contributing partonic channels $a b \to c X$ 
and $d \hat {\sigma}/ d \hat{t}$ is the associated partonic cross 
section. In these calculations $f_{a/p},f_{b/p}$ are our parton distributions
and $D_{\pi^0/c}$ is the pion fragmentation function.
Our calculations are done up to the NLO corrections, using the numerical code 
INCNLL of Ref. \cite{AVER} and for two different choices of fragmentation 
functions namely, BKK of Ref. \cite{binew95} and KKP of Ref. \cite{kniel00}, 
and we have checked that they give similar numerical results. We have compared
our predictions to two different data sets at $\sqrt{s} = 200\mbox{GeV}$
from PHENIX and STAR at RHIC-BNL. The results are shown in 
Figs.~\ref{fi:pi0phenix} and \ref{fi:pi0star} and the agreement is good, both 
in central rapidity region (PHENIX) and in the forward region (STAR). 
This energy is high enough to expect NLO QCD calculations to be valid
in a large rapidity region, which is not the case for lower energies 
\cite{BS0}.

\begin{figure}
\begin{center}
\leavevmode {\epsfysize=14.0cm \epsffile{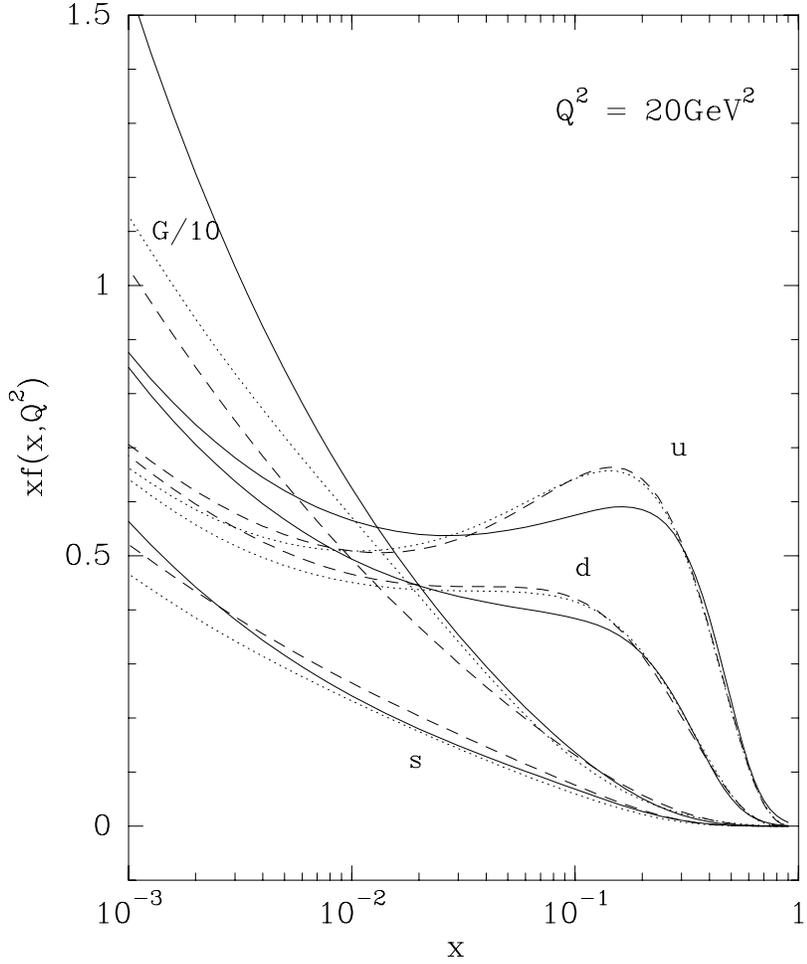}}
\end{center}
\vspace*{-5mm}
\caption[*]{\baselineskip 1pt
A comparison of the PDF at NLO from the statistical model (solid) 
with MRST2002
(dashed) \cite{MRST} and CTEQ6 (dotted) \cite{CTEQ}, for quarks $u, d, s$
and gluon at $Q^2 = 20\mbox{GeV}^2$.}
\label{fi:pdf}
\vspace*{-1.5ex}
\end{figure}

\clearpage
\newpage
\section{Concluding remarks}
We have shown that this simple approach of the statistical parton distributions
provides a good description of recent data on unpolarized and polarized DIS 
and on several hadronic processes. 
Since it involves only eight free parameters, we have tried to relate them to 
some specific properties of the parton distributions, but we do not
have yet a full understanding of their physical interpretation. It is 
important to stress that we have simultaneously the unpolarized and the 
polarized parton distributions, which is a unique situation. 
The main features of our distributions agree with other sets available in the 
literature, both in the unpolarized case \cite{MRST,CTEQ} see 
Fig.~\ref{fi:pdf} and in the polarized case \cite{GRSV, AAC, LSS}. We show in 
Fig.~\ref{fi:pdf}
a comparison with MRST and CTEQ, where one observes that the essential 
differences lie in the small $x$ region.
We have also identified some physical
observables and kinematic regions, where we can make definite predictions.
In particular, let us recall a slow decreasing behavior of 
$d(x)/u(x)$ for $x>0.6$, the fact that $\bar {d}(x)/\bar {u}(x)$ should 
remain larger than one for $x>0.3$, the signs $\Delta \bar {u}(x)>0$ and 
$\Delta \bar {d}(x)<0$ and our choice for $\Delta G(x)$. 
All these are real challenges and we look forward to new precise 
experimental data in the future.\\
\noindent
{\bf{Aknowledgments}}\\
F. Buccella wishes to thank the Centre de Physique Th\'eorique, where this
work was done, for warm hospitality and the Universit\'e de Provence for
financial support.
\newpage
\clearpage

\end{document}